\documentclass[letterpaper,showpacs,english,aps,preprint,nofootinbib]{revtex4}
\usepackage{amsmath}
\usepackage{graphicx}
\usepackage{amssymb}
\usepackage{color}
\makeatletter

\usepackage{babel}
\makeatother

\begin{document}
\begin{flushright}
MSUHEP-040818 \\
hep-ph/0408180
\end{flushright}

\newcommand{\BKMM}[2]{\left<\hat{#1}\!-\!|\hat{#2}-\right>}
\newcommand{\BKPP}[2]{\left<\hat{#1}\!+\!|\hat{#2}+\right>}
\newcommand{\BKM}[2]{\left<\hat{#1}\!-\!|\hat{#2}+\right>}
\newcommand{\BKP}[2]{\left<\hat{#1}\!+\!|\hat{#2}-\right>}
\newcommand{\BSKM}[3]{\left<\!\hat{#1}\!-\!|\not\!#2|\hat{#3}-\right>}
\newcommand{\BSKMMP}[3]{\left<\!\,\hat{#1}\!-\!|\not{\!#2}_{-}|\hat{#3}+\right>}
\newcommand{\BSKPMM}[3]{\left<\!\,\hat{#1}\!+\!|\not{\!#2}_{-}|\hat{#3}-\right>}
\newcommand{\BSKP}[3]{\left<\hat{#1}\!\!+\!|\!\not{\!#2}_{-}|\hat{#3}+\right>}
\newcommand{\BSSKM}[4]{\left<\hat{#1}\!-\!|\not{\!#2}_{+}\not{\!#3}_{-}|\hat{#4}+\right>}
\newcommand{\BSSKP}[4]{\left<\hat{#1}\!\!+\!|\not{\!#2}_{-}\not{\!#3}_{+}|\hat{#4}-\right>}
\newcommand{\BSSKMPMM}[4]{\left<\hat{#1}\!-\!|\not{\!#2}_{+}\not{\!#3}_{-}|\hat{#4}-\right>}
\newcommand{\BSSKPPMP}[4]{\left<\hat{#1}\!+\!|\not{\!#2}_{+}\not{\!#3}_{-}|\hat{#4}+\right>}
\newcommand{\BSSKPMPP}[4]{\left<\hat{#1}\!+\!|\not{\!#2}_{-}\not{\!#3}_{+}|#4+\right>}
\newcommand{\BSSSKM}[5]{\left<\hat{#1}\!-\!|\not{\!#2}_{+}\not{\!#3}_{-}\not{\!#4}_{+}|\hat{#5}-\right>}
\newcommand{\BSSSKP}[5]{\left<\hat{#1}\!\!+\!|\not{\!#2}_{-}\not{\!#3}_{+}\not{\!#4}_{-}|\hat{#5}+\right>}

\title{Single Top Quark Production and Decay at Next-to-leading Order in
Hadron Collision }

\author{Qing-Hong Cao}
\email{cao@pa.msu.edu}
\author{C.--P. Yuan}
\email{yuan@pa.msu.edu}

\affiliation{
\vspace*{2mm}
{Department of Physics $\&$ Astronomy, \\
Michigan State University, \\
East Lansing, MI 48824, USA.\\ }}

\vspace{0.15in}

\begin{abstract}
We present a calculation of the next-to-leading order QCD corrections,
with one-scale phase space slicing method, to single top quark production
and decay process $p\bar{p},pp\rightarrow t\bar{b}+X\rightarrow b\ell\nu\bar{b}+X$
at hadron colliders. Using the helicity amplitude method, the angular
correlation of the final state partons and the spin correlation of
the top quark are preserved. The effect of the top quark width is
also examined.
\end{abstract}


\maketitle

\section{Introduction }

At hadron colliders, the top quarks ($t$) are predominantly produced
in pairs via the strong interaction process $q{\bar{q}},\, gg\rightarrow t\bar{t}$.
Though it is possible to study the decay branching ratios of the top
quark in $t\bar{t}$ pairs, to test the coupling of top quark with
bottom quark ($b$) and $W$ gauge boson in hadron collisions, it is best to study the single-top
quark production. Compared to the top quark pair production, produced
by the interaction of the Quantum Chromodynamics (QCD), the single
top quark productions are through the electroweak interaction connecting
top quark to the down-type quarks, with amplitudes proportional to
the Cabibbo-Kabayaashi-Maskawa (CKM) matrix elements. Due to the nature
of left-handed charged weak current interaction, the top quark produced
via single-top processes is highly polarized. Furthermore, top quark
will decay via weak interaction before it has a chance to form a hadron,
so its polarization property can be studied from the angular distributions
of its decay particles. Hence, measuring the production rate of the
single-top event can directly probe the electroweak properties of
the top quark. For example, it can be used to measure the CKM matrix
element $V_{tb}$ and to test the $V-A$ structure of the top quark
charged-current weak interaction, or to probe CP violation 
effects~\cite{Yuan:1994fn,Atwood:1996pd,Bar-Shalom:1997si}. 
Besides of playing the role as a test of the Standard
Model (SM), the precision measurement of the single top quark events
has additional importance in searching for new physics, because the
charged-current top quark coupling ($W$-$t$-$b$) might be particularly
sensitive to certain new physics via new weak interactions or via loop
effects, and new production mechanism might also contribute to the single
top event rate~\cite{Kane:1991bg,Carlson:1994bg,Rizzo:1995uv,Malkawi:1996fs,
Tait:1996dv,Datta:1996gg,Li:1996ir,Simmons:1996ws,Whisnant:1997qu,
Baringer:1997wu,Li:1997qf,Tait:1997fe,Hikasa:1998wx,Han:1998tp,
He:1998ie,Boos:1999dd,Espriu:2001vj}. Furthermore, the
single-top event is also an important background to the search of
Higgs boson ($q\bar{q}'\rightarrow WH$ with $H\rightarrow b\bar{b}$)
at the Tevatron~\cite{Stange:1993ya,Stange:1994bb,Belyaev:1995gb} and 
other new physics search~\cite{Cao:2003tr}.

Because of the unique features of the single top quark physics, it
has been extensively studied in the 
literature~\cite{Tait:1997fe,Cortese:1991fw,Stelzer:1995mi,Mrenna:1997wp,
Willenbrock:1997nr,Stelzer:1998ni,Belyaev:1998dn,Smith:1996ij,
Stelzer:1997ns,Dawson:1984gx,Willenbrock:1986cr,Yuan:1989tc,
Ellis:1992yw,Carlson:1993dt,Heinson:1996zm,Bordes:1994ki,
Ladinsky:1990ut,Moretti:1997ng,Tait:1999cf,Belyaev:2000me,Tait:2000sh,Zhu:2001hw}.
There are three separate single top quark production processes of
interest at the hadron collider, which may be characterized by the
virtuality of the $W$ boson (with four momentum $q$) in the processes.
The s-channel process $q\bar{q}^{\prime}\rightarrow W^{*}\rightarrow t\bar{b}$
via a virtual s-channel $W$ boson involves a timelike $W$ boson,
$q^{2}>(m_{t}+m_{b})^{2}$, the t-channel process $qb\rightarrow q't$
(including $\bar{q}'b\rightarrow\bar{qt}$, also referred as $W$-gluon
fusion) involves a spacelike $W$ boson, $q^{2}<0$, and the $tW$
associated production process $bg\rightarrow tW^{-}$ involves an
on-shell $W$ boson, $q^{2}=m_{W}^{2}$. Therefore, these three single
top quark production mechanisms probe the charged-current interaction
in different $q^{2}$ regions and are thus complementary to each other.
Furthermore, they are sensitive to different new physics
effects~\cite{Tait:2000sh}, and should be separately studied. 

To improve the theoretical prediction on the single-top production rate,
a next-to-leading-order (NLO) correction, at the order of $\alpha_{s}$,
for the s- and t-channel processes has been carried out in Refs.~\cite{Smith:1996ij,Stelzer:1997ns,Bordes:1994ki}.
This is similar to the study of the $O(\alpha_{s})$ correction to
the top quark decay~\cite{Jezabek:1994zv}.
Although the above studies provide
the inclusive rate for single-top production, they cannot predict
the event topology of the single-top event, which is crucial to confront
the theory with experimental data in which some kinematical cuts are
necessary to detect such an event. For that, Refs.~\cite{Harris:2002md,Sullivan:2004ie}
have calculated the differential cross section for on-shell single
top quark production. However, NLO corrections to the top quark decay process
were not included, nor the effects of the top quark width were considered. 
Since the top quark production and decay do not occur in isolation from each
other, a theoretical study that includes both kinds of corrections
is needed. A complete NLO calculation should include contribution
from the production and the decay of the top quark, and the angular
correlation among the final state particles should be calculated to
analyze the polarization of the top quark. The $O(\alpha_{s})$ corrections 
to kinematic distributions may depend on the kinematic
cuts and on the jet algorithm that must be implemented in the experiments.
Therefore, it is necessary to obtain a fully differential calculation
that can be used to study the kinematics of the final state particles.

In this theory paper, we present a NLO QCD calculation with the one-scale
phase space slicing method, which treats consistently $O(\alpha_{s})$
corrections to both the production and the decay of the top quark in single top
events. Our approach can give not only the inclusive total cross section,
but also the various kinematical distributions of the final state particles, and
provide a study on the top quark polarization at the NLO. Furthermore,
since realistic kinematical cuts can be applied, our approach
allows the experimentalists to compare their results directly with
the theoretical predictions. In our study, we assume in all cases
leptonic decays of the $W$ boson (for the sake of definiteness, we
shall consider $W^{+}\rightarrow e^{+}\nu$; the lepton mass effects will be neglected
throughout this paper). The phenomenological discussions will be given
in our sequential papers~\cite{pheno-schan}.

The rest of this paper is organized as follows. In Sec. II, we outline
the method of our calculation. In Sec. III, we present the Born level
helicity amplitudes of the production and decay of single top quark.
In Sec IV, we present the NLO helicity amplitudes of the production
and decay of single top quark. The effective form factor approach
is adopted in the calculation to generalize the application of our
formalism to, for example, studying new physics effects. In Sec. V,
we use the phase space slicing (PSS) method to calculate the effective
form factors. To regularize divergencies in the calculation that involves the
$\gamma_5$ matrix, both the dimensional regularization (DREG)~\cite{'tHooft:1972fi}
 and the dimensional reduction (DRED)~\cite{Bern:2002zk} schemes
are examined and their difference is shown
in each individual form factor. In Sec. VI, we show how to assemble
all the components discussed above to enumerate the NLO differential
cross section of the single top quark. Finally, we give our conclusions
in Sec. VII.

\section{Outline of the calculation}

In this section we outline the method of our calculation whose details
shall be presented in the following sections.

\subsection{Narrow width approximation \label{sub:nwa}}

In this work, the narrow width approximation (NWA) is used to study
the production and decay of single top quark, in which the $O(\alpha_{s})$
corrections can be unambiguously assigned to either the single top quark
production process or the top quark decay process. A finite top width
will result in a new type of virtual NLO Feynman diagram in which a
gluon line is connected from the anti-bottom quark (of the
production process) to the bottom quark (of the decay process). Moreover,
there will
also be interference between the gluons emitted in the production
and the gluon emitted in the decay if the effects of finite top quark
width is considered. Those effects are nonfactorizable,
which are similar to the effects of QED radiative corrections to the scattering
process $e^{+}e^{-}\rightarrow W^{+}W^{-}\rightarrow4f$. It was shown that 
the nonfactorizable effects are small as long as the process is not near the 
threshold~\cite{Melnikov:1995fx,Beenakker:1997ir,Beenakker:1997bp}. 
This provides the motivation of using the NWA for this kind of 
calculation~\cite{Schmidt:1995mr,Macesanu:2001bj,Pittau:1996rp}.

\begin{figure}
\includegraphics[%
  scale=0.6]{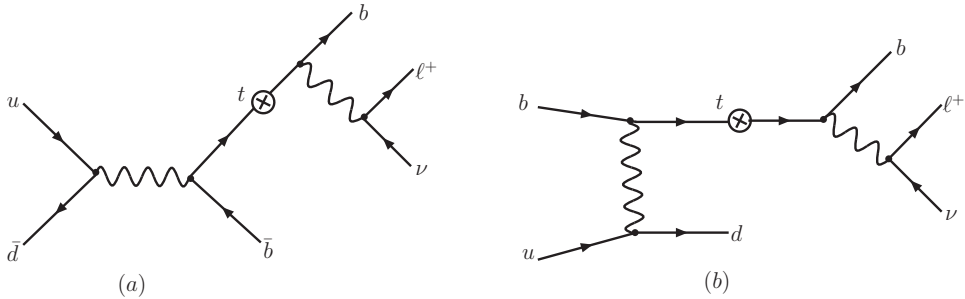}

\caption{Feynman diagrams of the Born level contribution to the production
and decay of single top quark. (a) s-channel (b) t-channel\label{fig:nwa-lo}}
\end{figure}

The single top quark can be produced through s-channel and t-channel
processes, as shown in Fig.~\ref{fig:nwa-lo}(a) and Fig.~\ref{fig:nwa-lo}(b),
respectively. Using the NWA, we decompose the Born level processes,
depicted in Fig.~\ref{fig:nwa-lo} as indicated by symbol $\otimes$,
into two parts: the top quark production and its sequential decay,
where both the production and decay matrices are separately gauge invariant. 
Making use of the polarization information of the top quark, we can apply the
NWA to correlate the top quark production with the top quark decay processes
by replacing the numerator of the top quark propagator ($\rlap/p_{t}+m_{t}$)
by $\sum_{\lambda_{t}=\pm}u^{\lambda_{t}}(t)\bar{u}^{\lambda_{t}}(t)$.
Here $u^{\lambda_{t}}(t)$ is the Dirac spinor of the top quark with
helicity $\lambda_{t}$, where $\lambda_{t}=+$ or $-$ for a right-handed
or left-handed top quark, respectively. Therefore, the scattering
amplitude of the single top quark production and decay processes can
be written as~\cite{Kane:1991bg}
\begin{equation}
\mathcal{M}=\sum_{\lambda_{t}=\pm}\mathcal{M}^{dec}(\lambda_{t})\mathcal{M}^{prod}(\lambda_{t}),\label{eq:DESNWA}
\end{equation}
where $\mathcal{M}(\lambda_{t})$ is the helicity amplitude and $\lambda_{t}$
is the helicity eigenvalue of the single top quark produced in the
intermediate state. The matrix element squared can be written as the
product of the production part and the decay part in the density matrix
formalism:
\begin{equation}
\left|\mathcal{M}\right|^{2}=\sum_{\lambda_{t},\lambda_{t}^{\prime}=\pm}\mathcal{A}_{\lambda_{t},\lambda_{t}^{\prime}}\mathcal{B}_{\lambda_{t},\lambda_{t}^{\prime}},\label{eq:density-matrix}\end{equation}
where \begin{eqnarray}
\mathcal{A}_{\lambda_{t},\lambda_{t}^{\prime}} & = & \mathcal{M}_{dec}^{\dagger}(\lambda_{t})\mathcal{M}_{dec}(\lambda_{t}^{\prime}),\label{eq:density-matrix-1}\\
\mathcal{B}_{\lambda_{t},\lambda_{t}^{\prime}} & = & \mathcal{M}_{prod}^{\dagger}(\lambda_{t})\mathcal{M}_{prod}(\lambda_{t}^{\prime}).\label{eq:density-matrix-2}\end{eqnarray}
In addition to the matrix elements, the phase space of the single top
quark processes can also be factorized into the top quark production
and decay for an on-shell top quark in NWA by writing the denominator
of the top quark propagator as\begin{equation}
\int dp^{2}\frac{1}{(p^{2}-m_{t}^{2})^{2}+m_{t}^{2}\Gamma_{t}^{2}}=\frac{\pi}{m_{t}\Gamma_{t}}.\label{eq:PSNWA}\end{equation}
When the matrix element is calculated using the fixed $m_t$ value,
it is the usual NWA method. 
In this case, the invariant mass of the top quark decay 
particles will be equal to $m_t$ (a fixed value) for all events.
Reconstructing the top quark invariant mass from its decay particles is an 
important experimental task at the Tevatron and the LHC, it is desirable
to have a theory calculation that would produce the invariant mass
distribution of the reconstructed top quark mass with a Breit-Wigner 
resonance shape to reflect the non-vanishing decay width of the top quark 
(for being an unstable resonance). For that, we introduce the 
 ``modified NWA'' method in our numerical calculation in
which we generate a Breit-Wigner distribution for the top quark invariant
mass in the phase space generator and then calculate the squared matrix 
element according to
Eq.~(\ref{eq:density-matrix}) with $m_t$ being the invariant mass 
generated by the phase space generator on the event-by-event basis.
In the limit that the total decay width of the top quark 
approaches to zero (i.e., much smaller
than the top quark mass), the production and the decay of top quark can
be factorized. Therefore, the S-matrix element for the production and
the decay processes are separately gauge invariant with any value of top
quark invariant mass. We find that the total event rate 
and the distributions of various kinematics variables 
(except the distribution of the reconstructed top quark invariant mass) 
calculated using the ``modified NWA'' method agree well with 
that using the NWA method. 
In the NWA method, the reconstructed top quark invariant mass
distribution is a
delta-function, i.e., taking a fixed value, while in the 
``modified NWA'' method, it is almost a Breit-Wigner distribution. 
The reason that the ``modified NWA'' method does not generate a perfect 
Breit-Wigner shape in the distribution of the top quark invariant mass 
is because the initial state 
parton luminosities (predominantly due to valence quarks) 
for the s-channel single-top process drop rapidly 
at the relevant Bjorken-$x$ range, where 
$\left\langle x\right\rangle \simeq \frac{m_t}{\sqrt{s}} \sim 0.1$. 

\subsection{Phase space slicing method\label{sub:Phase-space-slicing}}

When calculating NLO QCD corrections, one generally encounters
both ultraviolet (UV) and infrared (IR) (soft and collinear) divergencies.
The former divergencies can be removed by proper renormalization of
couplings and wave functions. We don't need to renormalize the couplings
in our calculation because the Born level couplings do not involve
QCD interaction. In order to handle the latter divergencies, one has
to consider both virtual and real corrections. The soft divergencies
will cancel according to the Kinoshita-Lee-Nauenberg (KLN) 
theorem~\cite{Kinoshita:1962ur,Lee:1964is},
but some collinear divergencies will remain uncanceled. In
the case of considering the initial state partons, one needs to absorb
additional collinear divergencies to define the NLO parton distribution
function (PDF) of the initial state partons. After that, all the infrared-safe
observables will be free of any singularities. To calculate the inclusive
production rate, one can use the dimensional regularization scheme to
regularize divergencies and adopt the modified minimal subtraction
($\overline{{\rm MS}}$) factorization scheme to obtain the total
rate. However, owing to the complicated phase space for multi-parton
configurations, analytic calculations are in practice impossible for
all but the simplest quantities. During the last few years, effective
numerical computational techniques have been developed to calculate
the fully differential cross section to NLO and above. There are,
broadly speaking, two types of algorithm used for NLO calculations:
the phase space slicing method, and the dipole subtraction 
method~\cite{Fabricius:1981sx,Kramer:1986mc,Giele:1991vf,Giele:1993dj,Keller:1998tf,
Ellis:1980wv,Catani:1996jh,Catani:1996vz,Harris:2001sx,Catani:2002hc}.
In this study, we use the phase space slicing method (PSS) with one
cutoff scale for which the universal crossing functions have been
derived in Ref.~\cite{Giele:1993dj}.
The advantage of this method
is that, after calculating the effective matrix elements with all
the partons in the final state, we can use the generalized crossing
property of the NLO matrix elements to calculate the corresponding
s-channel or t-channel matrix elements numerically without requiring
any further effort. The validity of this method is due to the property
that both the phase space and matrix element of the initial and final
state collinear radiation processes can be simultaneously factorized.
Below, we briefly review the general formalism of the NLO calculation
in PSS method with one cutoff scale.

The phase space slicing method with one cutoff scale introduces an
unphysical parameter $s_{min}$ to separate the real emission
correction phase space into two regions: (1) the resolved region in
which the amplitude has no divergencies and can be integrated numerically
by Monte Carlo method; (2) the unresolved region in which the amplitude
contains all the soft and collinear divergencies and can be integrated
out analytically. It should be emphasized that the notion of resolved/unresolved
partons is unrelated to the physical jet resolution criterium or to
any other relevant physical scale. In the massless case, a convenient
definition of the resolved region is given by the requirement $s_{ij}>s_{min}$
for all invariants $s_{ij}=(p_{i}+p_{j})^{2}$, where $p_{i}$ and
$p_{j}$ are the 4-momenta of partons $i$ and $j$, respectively.
For the massive quarks, we follow the definition in 
Ref.~\cite{Brandenburg:1997pu} to account for masses, 
but still use the terminology ``resolved''
and ``unresolved'' partons. In the regions with unresolved partons,
soft and collinear approximations of the matrix elements, which hold
exactly in the limit $s_{min}\rightarrow0$, are used. The necessary
integrations over the soft and collinear regions of phase space can
then be carried out analytically in $d=4-2\epsilon$ space-time dimensions.
One can thus isolate all the poles in $\epsilon$ and perform the
cancellation of the IR singularities between the real and virtual
contributions and absorb the leftover singularities into the parton
structure functions via the factorization procedure. After the above
procedure, one takes the limit $\epsilon\rightarrow0$. The contribution
from the sum of virtual and unresolved region corrections is finite
but $s_{min}$ dependent. Since the parameter $s_{min}$ is introduced
in the theoretical calculation for technical reasons only and is unrelated
to any physical quantity, the sum of all contributions (virtual, unresolved
and resolved corrections) must not depend on $s_{min}$. We note that the phase
space slicing method is only valid in the limit that $s_{{\rm min}}$
is small enough so that a given jet finding algorithm (or any infrared-safe
observable) can be consistently defined even after including the experimental
cuts.

\begin{figure}
\includegraphics[%
  scale=0.6]{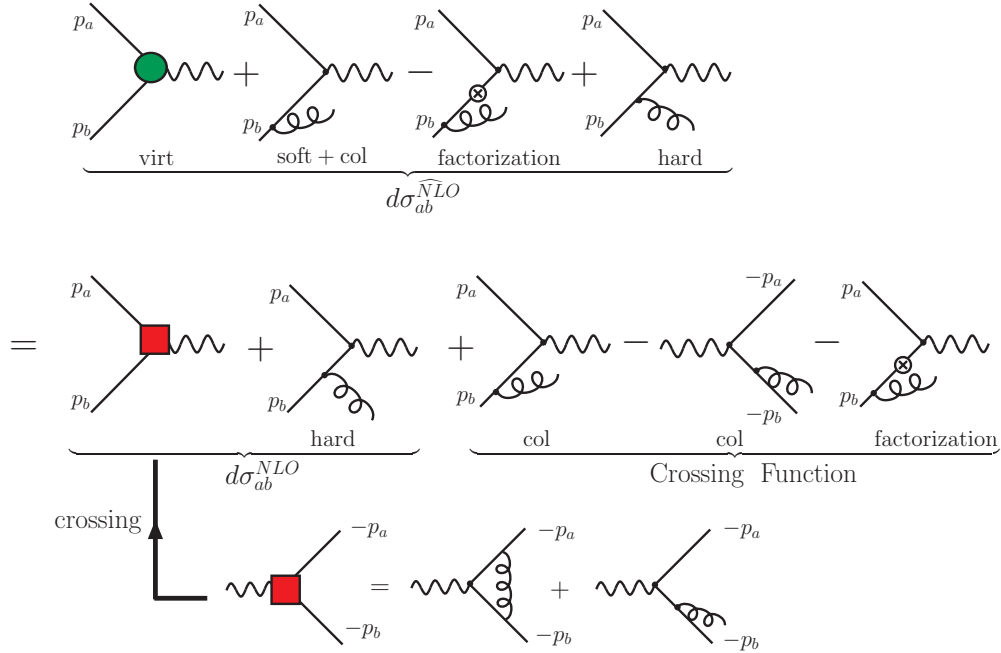}
\caption{Illustration of the PSS method with one cutoff scale to describe
the processes with initial state massless quarks. Here, only half
of the real emission diagrams is shown. In this paper, we assign
the particle's momentum such that the initial state particle's momentum
is incoming to the vertex while the final state particle's momentum
is outgoing.\label{fig:crossing}}
\end{figure}

In general, the conventional calculation of the NLO differential cross
section for a process with initial state hadrons $H_{1}$ and $H_{2}$
can be written as
\begin{equation}
d\sigma_{H_{1}H_{2}}^{NLO}=\sum_{a,b}\int dx_{1}dx_{2}f_{a}^{H_{1}}(x_{1},\mu_{F})f_{b}^{H_{2}}(x_{2},\mu_{F})
d\widehat{\sigma_{ab}^{NLO}}(x_{1,}x_{2},\mu_{R}),\label{eq:conventional_NLO}
\end{equation}
where $a$, $b$ denote parton flavors and $x_{1}$, $x_{2}$ are parton
momentum fractions. $f_{a}^{H}(x,\mu_{F})$ is the usual NLO parton
distribution function with the mass factorization scale $\mu_{F}$
and $d\widehat{\sigma_{ab}^{NLO}}(x_{1},x_{2},\mu_{R})$ is the NLO
hard scattering differential cross section with the renormalization
scale $\mu_{R}$. The diagrammatic demonstration of Eq.~(\ref{eq:conventional_NLO})
is shown in the upper part of Fig.~\ref{fig:crossing}. 

Unlike the conventional calculation method, the PSS method with
one cutoff scale will first cross the initial state partons into
the final state, including the virtual corrections and unresolved
real emission corrections. For example, to calculate the s-channel
single top quark production at the NLO, we first calculate the radiative
corrections to $W^{*}\rightarrow q\bar{q}'(g)$, as shown in the lower
part of the Fig.~\ref{fig:crossing}, in which we split the phase
space of the real emission corrections into the unresolved and resolved
region. After we integrate out the unresolved phase space region,
the net contribution of the virtual corrections and the real emission
corrections in the unresolved phase space is finite but theoretical
cutoff ($s_{min}$) dependent, and it can be written as a form factor
(denoted by the box in Fig.~\ref{fig:crossing}) of the Born level
vertex.

Secondly, we take the already calculated effective matrix elements
with all the partons in the final state and use the universal ``crossing
function'', which is the generalization of the crossing property
of the LO matrix elements to NLO, to calculate the corresponding matrix
elements numerically. Once we cross the needed partons back to the initial
state, the contributions from the unresolved collinear phase space
regions are different from those with all the partons in the final
state. These differences are included into the definition of the crossing
function (including the mass factorization effects), as shown in the
middle part of Fig.~\ref{fig:crossing}. In this paper, we only present
the explicit expressions of the crossing function. For the definition
and detailed derivation of the crossing function, we refer the reader
to Ref.~\cite{Giele:1993dj}.
After applying the mass factorization in
a chosen scheme, the $s_{min}$ dependent crossing functions for an initial state parton
$a$, which participates in the hard scattering processes, can be
written in the form:\begin{equation}
C_{a}^{{\rm scheme}}(x,\mu_{F},s_{min})=\left(\frac{N_{C}}{2\pi}\right)\left[A_{a}(x,\mu_{F})\log\left(\frac{s_{min}}{\mu_{F}}\right)+B_{a}^{{\rm scheme}}(x,\mu_{F})\right],\label{eq:crossing_def}\end{equation}
where\begin{eqnarray}
A_{a}(x,\mu_{F}) & = & \sum_{p}A_{p\rightarrow a}\left(x,\mu_{F}\right),\label{eq:crossing_def_a}\\
B_{a}^{{\rm scheme}}(x,\mu_{F}) & = & \sum_{p}B_{p\rightarrow a}^{{\rm scheme}}(x,\mu_{F}),\label{eq:crossing_def_b}\end{eqnarray}
and $N_{C}$ denotes the number of colors. The sum runs over $p=q,\bar{q},g$.
The functions $A$ and $B$ can be expressed as convolution integrals
over the parton distribution functions and their explicit forms are
shown in Appendix~\ref{sec:Crossing-Functions}. Although $A_{a}$
is scheme independent, $B_{a}$ does depend on the mass factorization
scheme, and so does the crossing function. 

After introducing the crossing function, we can write the NLO differential
cross section in the PSS method with one cutoff scale as
\begin{eqnarray}
d\sigma_{H_{1}H_{2}}^{NLO}& = & \sum_{a,b}\int dx_{1}dx_{2}\biggl\{f_{a}^{H_{1}}(x_{1},\mu_{F})f_{b}^{H_{2}}(x_{2},\mu_{F})d\sigma_{ab}^{NLO}(x_{1},x_{2},\mu_{R}) \label{eq:nlo-formalism-1} \\ 
 & +& \alpha_{s}(\mu_{R})\biggl[C_{a}^{H_{1}}(x_{1},\mu_{F})f_{b}^{H_{2}}(x_{2},\mu_{F})+f_{a}^{H_{1}}(x_{1},\mu_{F})C_{b}^{H_{2}}(x_{2},\mu_{F})\biggr]d\sigma_{ab}^{LO}(x_{1},x_{2})\biggr\}.\nonumber
\end{eqnarray}
Here $d\sigma_{ab}^{NLO}$ consists of the finite effective all-partons-in-the-final-state
matrix elements, in which partons $a$ and $b$ have simply been crossed
to the initial state, i.e. in which their momenta $-p_{a}$ and $-p_{b}$
have been replaced by $p_{a}$ and $p_{b}$, as shown in the Fig.~\ref{fig:crossing}.
The difference between $d\sigma_{ab}^{NLO}$ and $d\widehat{\sigma_{ab}^{NLO}}$
has been absorbed into the finite, universal crossing function $C_{a}^{H}(x,\mu_{F})$.
Defining an ``effective'' NLO parton distribution function $\mathcal{F}_{a}^{H}(x)$
as\begin{equation}
\mathcal{F}_{a}^{H}(x)=f_{a}^{H}(x,\mu_{F})+\alpha_{s}(\mu_{R})C_{a}^{H}(x,\mu_{F})+O(\alpha_{s}^{2}),\label{eq:effective-pdf}\end{equation}
we can rewrite Eq.~(\ref{eq:nlo-formalism-1}) in a simple form as

\begin{equation}
d\sigma_{H_{1}H_{2}}^{NLO}=\sum_{a,b}\int dx_{1}dx_{2}\mathcal{F}_{a}^{H_{1}}(x_{1})\mathcal{F}_{b}^{H_{2}}(x_{2})d\sigma_{ab}^{NLO}(x_{1},x_{2}).\label{eq:PSS-NLO-Formalism}\end{equation}

\subsection{$\gamma_{5}$ problem}

Because of the presence of the axial-vector current, a prescription
to handle the $\gamma_{5}$ matrices in $d$($=4-2\epsilon$) dimensions
has to be chosen. In this paper, we show the results of our calculations
using both the dimensional regularization (DREG) scheme ('t Hooft-Veltman
scheme~\cite{'tHooft:1972fi}) and the dimensional reduction (DRED) scheme (four
dimensional helicity scheme~\cite{Bern:2002zk}) to regulate the ultraviolet
and infrared divergencies presented in the NLO calculations. We note
that the results of form factors and the crossing function should
be done consistently in a given scheme. Except for the top quark
mass renormalization, we work in the $\overline{{\rm MS}}$ scheme
throughout the paper to perform the needed renormalization and factorization
procedures in order to calculate any ultraviolet and infrared finite
physical observable. To renormalize the top quark mass, we use the on-shell subtraction scheme.

\section{Leading order results of single top quark production and decay process\label{sec:Born-level-matrix}}

In this section we present the leading order results of the single
top quark processes. Using the density matrix method in the NWA, cf. Eq.~(\ref{eq:density-matrix}),
we factorize the s-channel and t-channel single top quark processes
(cf. Fig.~\ref{fig:nwa-lo}) into the top quark production and decay,
separately. To compute the amplitudes we use the spinor helicity  methods~\cite{Berends:1981rb,Kleiss:1985yh,Xu:1986xb,Gunion:1985vc,Mangano:1990by}
with the conventions as in Ref.~\cite{carlson}, and for completeness,
we briefly review the notation in Appendix~\ref{sec:Helicity-notation}.
We note that in some of Refs.~\cite{Berends:1981rb,Kleiss:1985yh,Xu:1986xb,Gunion:1985vc,Mangano:1990by}
 the phase conventions
do not correspond to the helicity convention utilized in this paper.
Below, we give the explicit Born level helicity amplitudes of the
single top quark production and decay, respectively.

\subsection{Helicity matrix elements of single top quark production}

The helicity amplitudes for the s-channel single top quark production
can be written as following:
\begin{eqnarray}
\mathcal{M}_{s}^{prod}(\lambda_{t}=+) & = & 2\BKPP{t}{\bar{d}}\BKP{u}{\bar{b}}\omega_{-}^{t},\label{eq:hel-s-tree-1}\\
\mathcal{M}_{s}^{prod}(\lambda_{t}=-) & = & 2\BKM{t}{\bar{d}}\BKP{u}{\bar{b}}\omega_{+}^{t},\label{eq:hel-s-tree-2}
\end{eqnarray}
where we have suppressed, for simplicity, the common factor ${\displaystyle \sqrt{2E_{u}}\sqrt{2E_{\bar{d}}}\sqrt{2E_{b}}}$,
the coupling constants ${\displaystyle \left(\frac{g}{\sqrt{2}}\right)^{2}}$
and the propagator ${\displaystyle \frac{1}{s-m_{W}^{2}}}$ with $s=(p_{u}+p_{\bar{d}})^{2}$.
Here, $g$ is the $SU(2)$ coupling constant, $m_{W}$ denotes the
mass of $W$-boson, and $\omega_{\pm}^{t}=\sqrt{E_{t}\pm\left|\vec{p_{t}}\right|}$,
where $E_{t}$ and $\vec{p_{t}}$ are the energy and momentum of the
top quark, respectively. The meaning of the bra ($<\mid$) and ket ($\mid>$) in the above
helicity amplitudes is summarized in Appendix~\ref{sec:Helicity-notation}.
We note that $\hat{u},\hat{\bar{d}},\hat{t}$ and $\hat{\bar{b}}$ within the bra 
and ket  denote the normalized three-momentum of the particle,
cf. Eq.~(\ref{eq:braket}). We did not write explicitly the helicity states of the other massless quarks
because only one set of the helicity states give a nonvanishing matrix element. For example, 
in this case the incoming $u$-quark is left-handed, $\bar{d}$ is right-handed and $\bar{b}$ is right-handed.
To calculate the squared matrix element, one also needs to include the proper spin and color factors which are not
explicitly shown in this paper.

For the t-channel single top quark production process, the helicity
amplitudes are given by
\begin{eqnarray}
\mathcal{M}_{t}^{prod}(\lambda_{t}=+) & = & 2\BKP{u}{b}\BKPP{t}{d}\omega_{-}^{t},\label{eq:hel-t-tree-1}\\
\mathcal{M}_{t}^{prod}(\lambda_{t}=-) & = & 2\BKP{u}{b}\BKM{t}{d}\omega_{+}^{t},\label{eq:hel-t-tree-2}
\end{eqnarray}
where we have also suppressed the common factor ${\displaystyle \sqrt{2E_{u}}\sqrt{2E_{d}}\sqrt{2E_{b}}}$,
the coupling constants ${\displaystyle \left(\frac{g}{\sqrt{2}}\right)^{2}}$,
and the propagator ${\displaystyle \frac{1}{t-m_{W}^{2}}}$ with $t=(p_{u}-p_{d})^{2}$.

\subsection{Helicity matrix elements of top quark decay}

For the top quark decay process, the helicity amplitude are given by
\begin{eqnarray}
\mathcal{M}^{dec}(\lambda_{t}=+) & = & -2\BKM{b^{\prime}}{\nu}\BKPP{e}{t}\omega_{-}^{t},\label{eq:hel-tdec-tree-1}\\
\mathcal{M}^{dec}(\lambda_{t}=-) & = & -2\BKM{b^{\prime}}{\nu}\BKP{e}{t}\omega_{+}^{t},\label{eq:hel-tdec-tree-2}
\end{eqnarray}
where we only consider the leptonic decay mode of $W$ boson and suppress,
for simplicity, the common factor ${\displaystyle \sqrt{2E_{e}}\sqrt{2E_{\nu}}\sqrt{2E_{b^{\prime}}}}$,
the coupling constants ${\displaystyle \left(\frac{g}{\sqrt{2}}\right)^{2}}$,
and the propagator ${\displaystyle \frac{1}{(p_{W}^{2}-m_{W}^{2})+im_{W}\Gamma_{W}}}$,
where $p_{W}$ and $\Gamma_{W}$ are the 4-momenta and the total decay
width of $W$-boson, respectively. All through out this paper we use $b^{\prime}$
to denote the bottom quark from top decay.

\section{NLO matrix elements of Single top quark production and decay processes\label{sec:NLO-matrix-element}}

Beyond the leading order, an additional gluon can be radiated from
the quark lines or appear as the initial parton in the single top
quark process . Since the single top quark can only be produced through
the electroweak interaction in the SM, we can further separate the
single top quark processes into smaller gauge invariant sets, even
at the NLO. Taking advantage of this property, in the first part of
this section we separate the s-channel and t-channel single top quark
processes into smaller gauge invariant sets of diagrams to organize
our calculations. As we pointed out in Sec.~\ref{sub:Phase-space-slicing},
NLO QCD corrections in the PSS method can be separated into two parts:
(I) the resolved real emission corrections and (II) the virtual correction
plus the unresolved real (soft+collinear) emission corrections, denoted
by {}``SCV''. After integrating out the virtual gluon and the unresolved
partons, the SCV corrections can be written as form factors multiplying
the Born level vertex. The form factors either modify the Born level
coupling or give rise to new Lorentz structure of $W$ coupling to
fermions. In the second part of this section, we will write down the
most general form factors of the single top quark processes and show
their contribution to the helicity amplitudes for both s-channel and
t-channel processes explicitly. It is worthwhile to mention that the
form factor formalism presented here can be easily extended to study
new physics models whose effects also show up as form factors.
The derivation of the form factors for single top quark production
and decay processes as predicted by the SM can be found in the second part
of this section. The resolved corrections are also calculated using helicity
amplitude method and the results are shown in the third part of this
section.

\subsection{Categorizing the single top quark processes\label{sub:categorizing-the-single}}

\begin{figure}
\includegraphics[%
  scale=0.6]{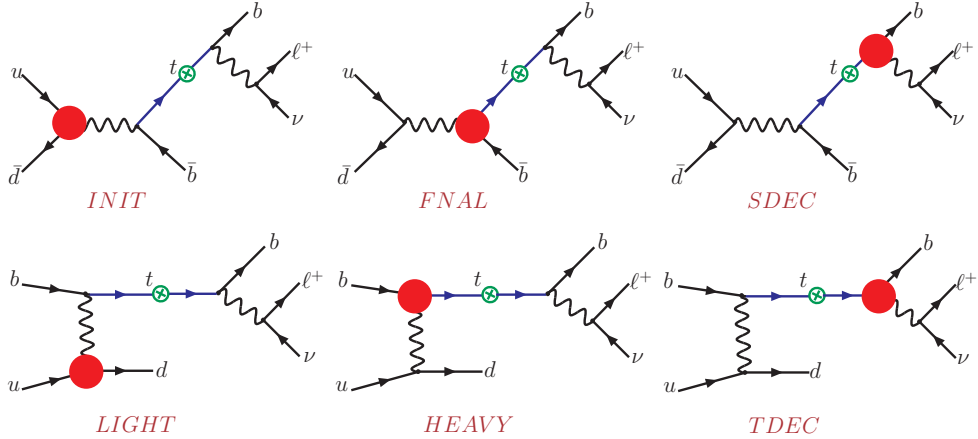}

\caption{The way we organize our calculations at the NLO. The blobs in the
diagrams denote the higher order QCD corrections, including both virtual
and real emission contributions.\label{fig:nlo}}
\end{figure}

Here, we separate the NLO s-channel and t-channel single top quark
processes into smaller gauge invariant sets of diagrams to organize
our calculations. The NLO s-channel diagrams consist of all the virtual
correction diagrams as well as the Feynman diagrams of the following
real correction processes: 
\begin{eqnarray}
 &  & q\bar{q}'\rightarrow W^{*}g\rightarrow\bar{b}gt(\rightarrow bW^{+}),\label{qqo}\\
 &  & qg\rightarrow W^{*}q'\rightarrow\bar{b}q't(\rightarrow bW^{+}),\label{qqp}\\
 &  & g\bar{q}'\rightarrow W^{*}\bar{q}\rightarrow\bar{b}\bar{q}t(\rightarrow bW^{+}),\label{qqt}\\
 &  & q\bar{q}'\rightarrow W^{*}\rightarrow\bar{b}gt(\rightarrow bW^{+}),\label{tb}\\
 &  & q\bar{q}'\rightarrow W^{*}\rightarrow\bar{b}t(\rightarrow bW^{+}g),\label{s-tdec}
\end{eqnarray}
where the gluon is connected only to ($q,q'$) lines in (\ref{qqo})-(\ref{qqt}),
and the gluon connected only to ($t,\bar{b}$) lines in (\ref{tb})
and the gluon connected only to ($t,b$) lines in (\ref{s-tdec}). We
note that diagrams (\ref{qqp}) and (\ref{qqt}) do not include those
in which the gluon line is connected to the final state $\bar{b}$
and $t$ line, for those are part of the NLO corrections to t-channel
process as shown in Eqs.~(\ref{qg}) and (\ref{qbarg}). To facilitate
the presentation of our calculation, we separate the s-channel higher
order QCD corrections (including both virtual and real corrections)
into the following three categories: 

\begin{itemize}
\item corrections to the initial state of the s-channel single top quark production
(INIT), in which the gluon is only connected to the initial state
light quark ($q,\bar{q}^{\prime}$) line,
\item corrections to the final state of the s-channel single top quark production
(FINAL), in which the gluon is only connected to the final state heavy
quark ($t,\bar{b}$) line of the single top quark production, 
\item corrections to the decay of the top quark (SDEC), in which the gluon
is connected to the heavy quark ($t,b$) line of the top quark decay.
\end{itemize}
The three types of corrections are illustrated in the upper part of Fig.~\ref{fig:nlo}, in which
the blobs represent the higher order QCD corrections. The explicit
real emission diagrams for the s-channel process can be found in Fig.~\ref{fig:real_s-chan}.

The NLO t-channel real correction processes for the top quark production
and decay are
\begin{eqnarray}
 &  & bq\rightarrow q'gt(\rightarrow bW^{+}),\label{qbtw}\\
 &  & b\bar{q}'\rightarrow\bar{q}gt(\rightarrow bW^{+}),\label{qbarbtw}\\
 &  & qg\rightarrow q'\bar{b}t(\rightarrow bW^{+}),\label{qg}\\
 &  & \bar{q}'g\rightarrow\bar{q}\bar{b}t(\rightarrow bW^{+}),\label{qbarg}\\
 &  & bg\rightarrow\bar{q}q't(\rightarrow bW^{+}),\label{bg}\\
 &  & bq\rightarrow q^{\prime}t(\rightarrow bW^{+}g),\label{t-topdec1}\\
 &  & b\bar{q}^{\prime}\rightarrow\bar{q}t(\rightarrow bW^{+}g).\label{t-topdec2}
\end{eqnarray}
 Here the gluon is connected to both ($q,q'$) lines and ($t,\bar{b}$)
lines in (\ref{qbtw}, \ref{qbarbtw}), but only to ($t,b$) lines in
(\ref{qg}, \ref{qbarg}). In (\ref{bg}), we restrict the gluon to
be connected only to ($q,q'$) lines. When the gluon in (\ref{bg})
is connected to ($t,b$) lines, it corresponds to the process $bg\rightarrow tW$
with $W\rightarrow\bar{q}q'$, therefore it is not included here.
As done in the s-channel case, we separate the t-channel NLO QCD corrections (including
both virtual and real corrections)(\ref{qbtw}-\ref{t-topdec2})
into three categories. As illustrated in the lower part of Fig.~\ref{fig:nlo}, they are:

\begin{itemize}
\item the one in which the gluon is connected to the light quark ($q,q'$) lines (LIGHT), 
\item the one in which the gluon is connected to the heavy quark ($t,\bar{b}$) lines (HEAVY),
\item the one in which the gluon is radiated from the heavy quark ($t,b$) lines of 
the on-shell top quark decay processes (TDEC), 
\end{itemize}
The explicit real emission diagrams for the t-channel process can be found in Fig.~\ref{fig:real_t-chan}.

\subsection{Form factor formalism for SCV corrections\label{sub:Form-factor-formalism}}

Below, we present the form factor formalism for each category defined
in Sec.~\ref{sub:categorizing-the-single}, after including both
the virtual and unresolved real corrections.

\subsubsection{INIT form factors}

Higher order QCD corrections to the diagrams labeled as INIT in Fig.~\ref{fig:nlo}
do not change the Lorentz structure of the $W^{*}-u-d$ coupling,
therefore the most general form of the initial state contribution
can be rewritten as
\begin{equation}
\frac{ig}{\sqrt{2}}\gamma^{\mu}P_{L}I_{L},\label{eq:form-s-init}
\end{equation}
where $I_{L}$ denotes the effective form factor that includes the
higher order corrections. Denoting the helicity amplitude as $\mathcal{M}_{INIT}(\lambda_{t})$
with top quark helicity $\lambda_{t}=\pm1$ and suppressing, for simplicity,
the common factor $\sqrt{2E_{u}}\sqrt{2E_{d}}\sqrt{2E_{\bar{{b}}}}$,
the coupling factors $\left({\displaystyle \frac{g}{\sqrt{2}}}\right)^{2}$
and the propagators ${\displaystyle \frac{1}{s-m_{W}^{2}}}$ with
$s=(p_{u}+p_{\bar{d}})^{2}$, we obtain the helicity amplitudes which
include higher order corrections to the initial state of the s-channel
single top quark production as following:
\begin{eqnarray}
\mathcal{M}_{INIT}(+) & = & 2I_{L}\BKPP{t}{\bar{d}}\BKP{u}{\bar{{b}}}\omega_{-}^{t},\label{eq:hel-s-init-scv-1}\\
\mathcal{M}_{INIT}(-) & = & 2I_{L}\BKM{t}{\bar{d}}\BKP{u}{\bar{{b}}}\omega_{+}^{t},\label{eq:hel-s-init-scv-2}
\end{eqnarray}
where $\omega_{\pm}^{t}=\sqrt{E_{t}\pm|\overrightarrow{p_{t}}|}$,
cf. Appendix~\ref{sec:Helicity-notation}.

Needless to say, when calculating the scattering amplitude of the single top quark production
and decay process, cf. Eq.~(\ref{eq:DESNWA}), up to the NLO, the decay matrix elements in this
case is taken to be the Born level ones as given in Eqs.~(\ref{eq:hel-s-tree-1})
and (\ref{eq:hel-s-tree-2}).

\subsubsection{FINAL form factors}

In the limit that the bottom quark mass is taken to be zero~%
\footnote{We take the bottom quark mass to be zero throughout our calculation
because $(m_{b}/m_{t})^{2}$ can be ignored numerically. Strictly
speaking, $\alpha_{s}\ln(m_{b})$ terms have been included in the
definition of NLO PDF.%
}, the most general $W^{\star}-t-b$ coupling, labeled as FINAL in
Fig.~\ref{fig:nlo}, is 
\begin{equation}
\frac{ig}{\sqrt{2}}\left\{ \gamma_{\mu}(F_{1}^{L*}P_{L}+F_{1}^{R*}P_{R})-\frac{(t_{\mu}-\bar{b}_{\mu})}{m_{W}}(F_{2}^{R*}P_{L}+F_{2}^{L*}P_{R})\right\} ,\label{eq:form-s-fnal-1}
\end{equation}
where the asterisk in the superscript of the form factors 
indicates taking its complex conjugate.
This is different from the coupling in Eq.~(\ref{eq:form-s-init})
because the top quark mass is kept in the calculation,and only the
bottom quark mass is taken to be zero. Because the charged current
interacts with massless quarks in the initial state, one can use the
on-shell condition of the massless initial state quarks to rewrite
Eq.~(\ref{eq:form-s-fnal-1}) as 
\begin{equation}
\frac{ig}{\sqrt{2}}\left\{ \gamma_{\mu}(F_{1}^{L*}P_{L}+F_{1}^{R*}P_{R})+\bar{b}_{\mu}(F_{2}^{R*}P_{L}+F_{2}^{L*}P_{R})\right\} ,\label{eq:form-s-fnal}
\end{equation}
where the $m_{W}$ has been absorbed into form factors $F_{2}^{R*}$
and $F_{2}^{L*}$. Denoting the helicity amplitude as $\mathcal{M}_{FINAL}(\lambda_{\bar{b}},\lambda_{t})$,
we obtain the helicity amplitudes which include higher order corrections
to the s-channel single top quark production as following:
\begin{eqnarray}
\mathcal{M}_{FINAL}(-,-) & = & 2F_{1}^{L*}\BKM{t}{\bar{d}}\BKP{u}{\bar{b}}\omega_{+}^{t}+F_{2}^{R*}\BSKM{\bar{d}}{\bar{b}}{u}\BKMM{t}{\bar{b}}\omega_{-}^{t},\label{eq:hel-s-fnal-scv-1}\\
\mathcal{M}_{FINAL}(+,-) & = & 2F_{1}^{R*}\BKMM{t}{u}\BKM{\bar{d}}{\bar{b}}\omega_{-}^{t}+F_{2}^{L*}\BSKM{\bar{d}}{\bar{b}}{u}\BKM{t}{\bar{b}}\omega_{+}^{t},\label{eq:hel-s-fnal-scv-2}\\
\mathcal{M}_{FINAL}(-,+) & = & 2F_{1}^{L*}\BKPP{t}{\bar{d}}\BKP{u}{\bar{b}}\omega_{-}^{t}+F_{2}^{R*}\BSKM{\bar{d}}{\bar{b}}{u}\BKP{t}{\bar{b}}\omega_{+}^{t},\label{eq:hel-s-fnal-scv-3}\\
\mathcal{M}_{FINAL}(+,+) & = & 2F_{1}^{R*}\BKP{t}{u}\BKM{\bar{d}}{\bar{b}}\omega_{+}^{t}+F_{2}^{L*}\BSKM{\bar{d}}{\bar{b}}{u}\BKPP{t}{\bar{b}}\omega_{-}^{t}.\label{eq:hel-s-fnal-scv-4}
\end{eqnarray}
As before, we have suppressed the common factor $\sqrt{2E_{u}}\sqrt{2E_{d}}\sqrt{2E_{\bar{{b}}}}$,
the coupling factors $\left({\displaystyle \frac{g}{\sqrt{2}}}\right)^{2}$,
and the propagators ${\displaystyle \frac{1}{s-m_{W}^{2}}}$ with
$s=(p_{u}+p_{\bar{d}})^{2}$.

\subsubsection{LIGHT form factors}

The effective form factor for $u-W^{*}-d$, labeled as LIGHT in Fig.~\ref{fig:nlo},
takes the exact same form as $W^{*}-u-d$ in Eq.~(\ref{eq:form-s-init}).
Hence, the helicity amplitudes $\mathcal{M}_{LIGHT}(\lambda_{t})$
for the t-channel single top quark production are given as follows:
\begin{eqnarray}
\mathcal{M}_{LIGHT}(+) & = & 2L_{L}\BKPP{t}{d}\BKP{u}{b}\omega_{-}^{t},\label{eq:hel-t-light-scv-1}\\
\mathcal{M}_{LIGHT}(-) & = & 2L_{L}\BKM{t}{d}\BKP{u}{b}\omega_{+}^{t},\label{eq:hel-t-light-scv-2}
\end{eqnarray}
where $L_{L}$ is the effective coupling induced by higher order corrections.
Again, we have suppressed the common factor $\sqrt{2E_{u}}\sqrt{2E_{d}}\sqrt{2E_{b}}$,
the coupling factors $\left({\displaystyle \frac{g}{\sqrt{2}}}\right)^{2}$
and the propagators ${\displaystyle \frac{1}{t-m_{W}^{2}}}$ with
$t=(p_{u}-p_{d})^{2}$.

\subsubsection{HEAVY form factors}

The effective form factor for $b-W^{*}-t$, labeled as HEAVY in Fig.~\ref{fig:nlo},
takes the exact same form as $W^{*}-t-b$ in Eq.~(\ref{eq:form-s-fnal}).
Hence, the helicity amplitudes $\mathcal{M}_{HEAVY}(\lambda_{b},\lambda_{t})$
for the t-channel single top quark production are given as follows:\begin{eqnarray}
\mathcal{M}_{HEAVY}(-,-) & = & 2H_{1}^{L*}\BKM{t}{d}\BKP{u}{b}\omega_{+}^{t}-H_{2}^{R*}\BSKM{d}{b}{u}\BKMM{t}{b}\omega_{-}^{t},\label{eq:hel-t-heavy-scv-1}\\
\mathcal{M}_{HEAVY}(+,-) & = & 2H_{1}^{R*}\BKMM{t}{u}\BKM{d}{b}\omega_{-}^{t}-H_{2}^{L*}\BSKM{d}{b}{u}\BKM{t}{b}\omega_{+}^{t},\label{eq:hel-t-heavy-scv-2}\\
\mathcal{M}_{HEAVY}(-,+) & = & 2H_{1}^{L*}\BKPP{t}{d}\BKP{u}{b}\omega_{-}^{t}-H_{2}^{R*}\BSKM{d}{b}{u}\BKP{t}{b}\omega_{+}^{t},\label{eq:hel-t-heavy-scv-3}\\
\mathcal{M}_{HEAVY}(+,+) & = & 2H_{1}^{R*}\BKP{t}{u}\BKM{d}{b}\omega_{+}^{t}-H_{2}^{L*}\BSKM{d}{b}{u}\BKPP{t}{b}\omega_{-}^{t},\label{eq:hel-t-heavy-scv-4}\end{eqnarray}
where $H_{1,2}^{L,R}$ denote the effective couplings induced by higher
order corrections. Here, we have suppressed the common factor $\sqrt{2E_{u}}\sqrt{2E_{d}}\sqrt{2E_{b}}$,
the coupling factors $\left({\displaystyle \frac{g}{\sqrt{2}}}\right)^{2}$
and the propagators ${\displaystyle \frac{1}{t-m_{W}^{2}}}$ with
$t=(p_{u}-p_{d})^{2}$.

\subsubsection{Top quark decay form factors}

The most general $t-b-W$ coupling, labeled as DEC for both s-channel
and t-channel processes, is
\begin{equation}
\frac{ig}{\sqrt{2}}\left\{ \gamma_{\mu}(D_{1}^{L}P_{L}+D_{1}^{R}P_{R})-b_{\mu}^{\prime}(D_{2}^{R}P_{R}+D_{2}^{L}P_{L})\right\} ,\label{eq:form-topdec}
\end{equation}
where $D_{1,2}^{L,R}$ denote the form factors which include higher
order QCD corrections. Denoting the helicity amplitude as $\mathcal{M}_{DEC}(\lambda_{t},\lambda_{b^{\prime}})$,
we obtain the helicity amplitudes which include higher order corrections
to single top quark decay process as following:\begin{eqnarray}
\mathcal{M}_{DEC}(-,-) & = & -2D_{1}^{L}\BKM{b^{\prime}}{\nu}\BKP{e}{t}\omega_{+}^{t}+D_{2}^{R}\BSKM{\nu}{b^{\prime}}{e}\BKMM{b^{\prime}}{t}\omega_{-}^{t},\label{eq:hel-topdec-scv-1}\\
\mathcal{M}_{DEC}(+,-) & = & -2D_{1}^{L}\BKM{b^{\prime}}{\nu}\BKPP{e}{t}\omega_{-}^{t}+D_{2}^{R}\BSKM{\nu}{b^{\prime}}{e}\BKM{b^{\prime}}{t}\omega_{+}^{t},\label{eq:hel-topdec-scv-2}\\
\mathcal{M}_{DEC}(-,+) & = & -2D_{1}^{R}\BKP{b^{\prime}}{e}\BKMM{\nu}{t}\omega_{-}^{t}+D_{2}^{L}\BSKM{\nu}{b^{\prime}}{e}\BKP{b^{\prime}}{t}\omega_{+}^{t},\label{eq:hel-topdec-scv-3}\\
\mathcal{M}_{DEC}(+,+) & = & -2D_{1}^{R}\BKP{b^{\prime}}{e}\BKM{\nu}{t}\omega_{+}^{t}+D_{2}^{L}\BSKM{\nu}{b^{\prime}}{e}\BKPP{b^{\prime}}{t}\omega_{-}^{t},\label{eq:hel-topdec-scv-4}\end{eqnarray}
where we ignore the common factor ${\displaystyle \sqrt{2E_{e}}\sqrt{2E_{\nu}}\sqrt{2E_{b^{\prime}}}}$,
the coupling factors $\left({\displaystyle \frac{g}{\sqrt{2}}}\right)^{2}$
and the propagator ${\displaystyle \frac{1}{(p_{W}^{2}-m_{W}^{2})+im_{W}\Gamma_{W}}}$
with $p_{W}=p_{e^{+}}+p_{\nu}$.

\vspace {3mm}

The category SDEC (or TDEC) in Fig.~\ref{fig:nlo} is obtained by
convoluting the s-channel (or t-channel) Born level helicity amplitudes,
cf. Eqs.~(\ref{eq:hel-s-tree-1}) and (\ref{eq:hel-s-tree-2}) (or 
Eqs.~(\ref{eq:hel-t-tree-1}) and (\ref{eq:hel-t-tree-2})), with
the corresponding DEC amplitudes listed above.

\subsection{Helicity amplitudes of resolved contributions \label{sub:Resolved-Contributions}}

Here, we present the helicity amplitudes of resolved corrections for
each category defined in Sec.~\ref{sub:categorizing-the-single}.

\begin{figure}
\includegraphics[%
  scale=0.6]{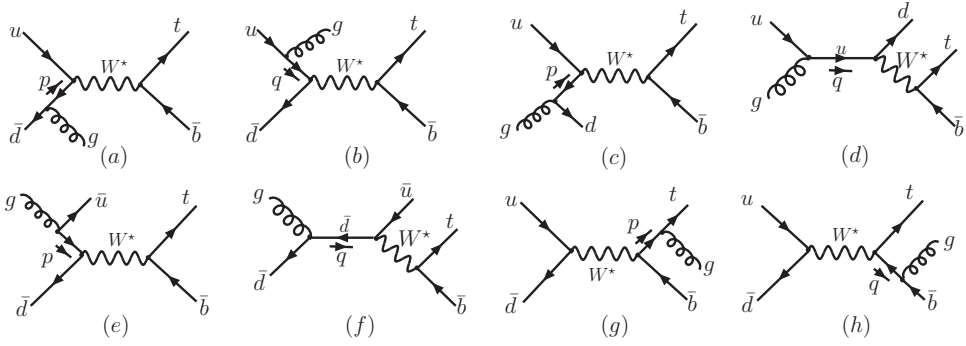}

\caption{Feynman diagrams of the real emission corrections to s-channel single
top quark production.\label{fig:real_s-chan}}
\end{figure}

\subsubsection{NLO corrections to INIT}

The Feynman diagrams of the initial state real emission corrections
are shown in Figs.~\ref{fig:real_s-chan}(a)-(f). At the NLO, a hard
gluon can be radiated from the initial state quark line, or a quark can
be generated from the gluon splitting. We separate the NLO INIT real
emission corrections into three categories:
\begin{eqnarray*}
{\rm INI-A} & : & q\bar{q}^{\prime}\rightarrow W^{*}g\rightarrow t\bar{b}g,\qquad{\rm including\,\,(a)\,\, and\,\,(b)},\\
{\rm INI-B} & : & qg\rightarrow W^{*}q^{\prime}\rightarrow t\bar{b}q^{\prime},\quad\,\,\,\,\,{\rm including\,\,(c)\,\, and\,\,(d),}\\
{\rm INI-C} & : & \bar{q}^{\prime}g\rightarrow W^{*}\bar{q}\rightarrow t\bar{b}\bar{q},\qquad{\rm including\,\,(e)\,\, and\,\,(f),}
\end{eqnarray*}
which are separately gauge invariant. Denoting the helicity amplitude
as $\mathcal{M}_{INI}^{A,B,C}(\lambda_{t})$, we calculate the helicity
amplitudes for a given helicity state ($\lambda_{t}$) of the top
quark, which are listed as follows. 

The helicity amplitudes for INI~A are :\begin{eqnarray}
\mathcal{M}_{INI}^{A}(+) & = & 2\,\omega_{-}^{t}\left\{ -\frac{\BKMM{\bar{d}}{t}\BSSKP{\bar{b}}{p}{\varepsilon^{*}}{u}}{p^{2}}+\frac{\BKP{\bar{b}}{u}\BSSKMPMM{\bar{d}}{\varepsilon^{*}}{q}{u}}{q^{2}}\right\} ,\label{eq:hel-s-inia-1}\\
\mathcal{M}_{INI}^{A}(-) & = & 2\,\omega_{+}^{t}\left\{ \frac{\BKM{\bar{d}}{t}\BSSKP{\bar{b}}{q}{\varepsilon^{*}}{u}}{p^{2}}-\frac{\BKP{\bar{b}}{u}\BSSKM{\bar{d}}{\varepsilon^{*}}{q}{t}}{q^{2}}\right\} ,\label{eq:hel-s-inia-2}\end{eqnarray}
with $p=p_{u}-p_{g}$ and $q=p_{\bar{d}}-p_{g}$. 

The helicity amplitudes for INI~B are :\begin{eqnarray}
\mathcal{M}_{INI}^{B}(+) & = & 2\,\omega_{-}^{t}\left\{ -\frac{\BKP{\bar{b}}{u}\BSSKMPMM{d}{\varepsilon}{p}{t}}{p^{2}}+\frac{\BKMM{d}{t}\BSSKP{\bar{b}}{q}{\varepsilon}{u}}{q^{2}}\right\} ,\label{eq:hel-s-inib-1}\\
\mathcal{M}_{INI}^{B}(-) & = & 2\,\omega_{+}^{t}\left\{ -\frac{\BKP{\bar{b}}{u}\BSSKM{\bar{d}}{\varepsilon}{p}{t}}{p^{2}}+\frac{\BKM{d}{t}\BSSKP{\bar{b}}{q}{\varepsilon}{u}}{q^{2}}\right\} ,\label{eq:hel-s-inib-2}\end{eqnarray}
with $p=p_{g}-p_{d}$ and $q=p_{g}+p_{u}$.

The helicity amplitudes for INI~C are :\begin{eqnarray}
\mathcal{M}_{INI}^{C}(+) & = & 2\,\omega_{-}^{t}\left\{ -\frac{\BKP{\bar{b}}{\bar{u}}\BSSKMPMM{\bar{d}}{\varepsilon}{q}{t}}{q^{2}}+\frac{\BKMM{\bar{d}}{t}\BSSKP{\bar{b}}{p}{\varepsilon}{\bar{u}}}{p^{2}}\right\} ,\label{eq:hel-s-inic-1}\\
\mathcal{M}_{INI}^{C}(-) & = & 2\,\omega_{+}^{t}\left\{ -\frac{\BKP{\bar{b}}{\bar{u}}\BSSKM{\bar{d}}{\varepsilon}{q}{t}}{q^{2}}+\frac{\BKM{\bar{d}}{t}\BSSKP{\bar{b}}{p}{\varepsilon}{\bar{u}}}{p^{2}}\right\} ,\label{eq:hel-s-inic-2}\end{eqnarray}
with $p=p_{g}-p_{\bar{u}}$ and $q=p_{g}+p_{\bar{d}}$.

Again in all the above equations, we have suppressed the common factor
$\sqrt{2E_{u}}\sqrt{2E_{\bar{d}}}\sqrt{2E_{\bar{b}}}$, the coupling
factors ${\displaystyle g_{s}\left(\frac{g}{\sqrt{2}}\right)^{2}}$,
and the propagator ${\displaystyle \frac{1}{p_{W}^{2}-m_{W}^{2}+im_{W}\Gamma_{W}}}$
with $p_{W}=p_{t}+p_{\bar{b}}$. Here, $g_{s}$ is the coupling constant
of the strong interaction.

\subsubsection{NLO corrections to FINAL}

The Feynman diagrams for NLO real emission corrections to the final
state of s-channel top quark production process are shown in the Fig.~\ref{fig:real_s-chan}(g)
and (h). Denoting the helicity amplitude as $\mathcal{M}_{FINAL}(\lambda_{t})$,
then\begin{eqnarray}
\mathcal{M}_{FINAL}(+) & = & 2\,\omega_{-}^{t}\frac{\BKP{u}{\bar{b}}\BSSKPPMP{t}{\varepsilon^{*}}{p}{\bar{d}}}{p^{2}-m_{t}^{2}}+2\, m_{t}\:\omega_{+}^{t}\frac{\BKP{u}{\bar{b}}\BSKP{t}{\varepsilon^{*}}{\bar{d}}}{p^{2}-m_{t}^{2}}\nonumber \\
 & - & 2\,\omega_{-}^{t}\frac{\BKPP{t}{\bar{d}}\BSSKP{u}{q}{\varepsilon^{*}}{\bar{b}}}{q^{2}},\label{eq:hel-s-fnal-res-1}\\
\mathcal{M}_{FINAL}(-) & = & 2\,\omega_{+}^{t}\frac{\BKP{u}{\bar{b}}\BSSKM{t}{\varepsilon^{*}}{p}{\bar{d}}}{p^{2}-m_{t}^{2}}+2\, m_{t}\,\omega_{-}^{t}\frac{\BKP{u}{\bar{b}}\BSKMMP{t}{\varepsilon^{*}}{\bar{d}}}{p^{2}-m_{t}^{2}}\nonumber \\
 & - & 2\,\omega_{+}^{t}\frac{\BKM{t}{\bar{d}}\BSSKP{u}{q}{\varepsilon^{*}}{\bar{b}}}{q^{2}},\label{eq:hel-s-fnal-res-2}\end{eqnarray}
with $p=p_{g}+p_{t}$ and $q=p_{g}+p_{\bar{b}}$. We again suppressed
the common factor $\sqrt{2E_{u}}\sqrt{2E_{\bar{d}}}\sqrt{2E_{\bar{b}}}$,
the coupling constants ${\displaystyle g_{s}\left(\frac{g}{\sqrt{2}}\right)^{2}}$,
and the $W$ boson propagator ${\displaystyle \frac{1}{p_{W}^{2}-m_{W}^{2}+im_{W}\Gamma_{W}}}$
with $p_{W}=p_{u}+p_{\bar{d}}$.

\begin{figure}
\includegraphics[%
  scale=0.6]{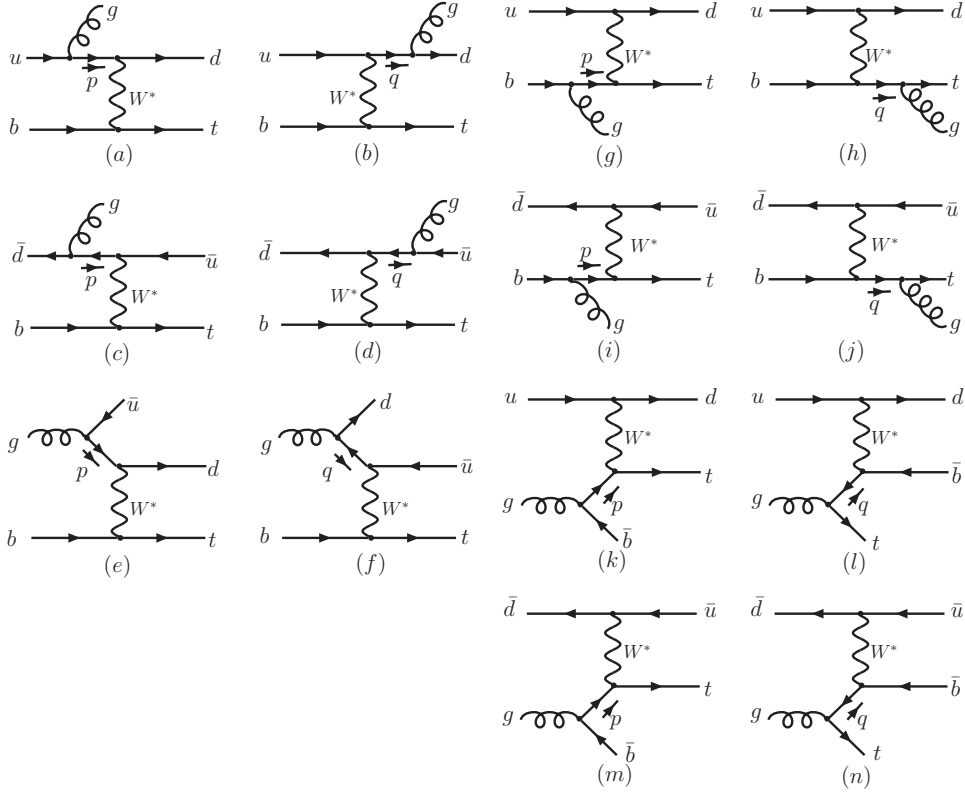}

\caption{Feynman diagrams of the real emission corrections to the t-channel
single top quark production\label{fig:real_t-chan}}
\end{figure}

\subsubsection{NLO corrections to LIGHT}

The Feynman diagrams, that generate real emission contributions through
coupling a gluon to the light quark lines, are shown in Figs.~\ref{fig:real_t-chan}(a)
to (f). To facilitate our calculations, we separate the NLO LIGHT
real emission corrections into the following three categories: \begin{eqnarray*}
{\rm LIGHT-A} & : & bq\rightarrow q'gt,\qquad{\rm including\,\,(a)\,\, and\,\,(b),}\\
{\rm LIGHT-B} & : & b\bar{q}'\rightarrow\bar{q}gt,\qquad{\rm including\,\,(c)\,\, and\,\,(d),}\\
{\rm LIGHT-C} & : & bg\rightarrow\bar{q}q't.\qquad{\rm \, including\,\,(e)\,\, and\,\,(f).}\end{eqnarray*}
Denoting the helicity amplitude as $\mathcal{M}_{LIGHT}^{A,B,C}(\lambda_{t})$,
then the helicity amplitudes for LIGHT-A are:\begin{eqnarray}
\mathcal{M}_{LIGHT}^{A}(+) & = & -2\omega_{-}^{t}\left\{ \frac{\BKMM{d}{t}\BSSKP{b}{p}{\varepsilon^{*}}{u}}{p^{2}}+\frac{\BKP{b}{u}\BSSKMPMM{d}{\varepsilon^{*}}{q}{t}}{q^{2}}\right\} ,\label{eq:hel-t-lighta-res-1}\\
\mathcal{M}_{LIGHT}^{B}(-) & = & \,\,\,\,2\omega_{+}^{t}\left\{ \frac{\BKM{d}{t}\BSSKP{b}{p}{\varepsilon^{*}}{u}}{p^{2}}+\frac{\BKP{b}{u}\BSSKM{d}{\varepsilon^{*}}{q}{t}}{q^{2}}\right\} ,\label{eq:hel-t-lighta-res-2}\end{eqnarray}
with $p=p_{u}-p_{g}$ and $q=p_{d}+p_{g}$.

The helicity amplitudes for LIGHT-B are:\begin{eqnarray}
\mathcal{M}_{LIGHT}^{B}(+) & = & \quad2\omega_{-}^{t}\left\{ \frac{\BKP{b}{\bar{u}}\BSSKMPMM{\bar{d}}{\varepsilon^{*}}{p}{t}}{p^{2}}+\frac{\BKMM{\bar{d}}{t}\BSSKP{b}{q}{\varepsilon^{*}}{\bar{u}}}{q^{2}}\right\} ,\label{eq:hel-t-lightb-res-1}\\
\mathcal{M}_{LIGHT}^{B}(-) & = & -2\omega_{+}^{t}\left\{ \frac{\BKP{b}{\bar{u}}\BSSKM{\bar{d}}{\varepsilon^{*}}{p}{t}}{p^{2}}+\frac{\BKM{\bar{d}}{t}\BSSKP{b}{q}{\varepsilon^{*}}{\bar{u}}}{k^{2}}\right\} ,\label{eq:hel-t-lightb-res-2}\end{eqnarray}
with $p=p_{\bar{d}}-p_{g}$ and $q=p_{\bar{u}}+p_{g}$.

The helicity amplitudes for LIGHT-C are:\begin{eqnarray}
\mathcal{M}_{LIGHT}^{C}(+) & = & -2\omega_{-}^{t}\left\{ -\frac{\BKMM{d}{t}\BSSKP{b}{p}{\varepsilon}{\bar{u}}}{p^{2}}+\frac{\BKP{b}{\bar{u}}\BSSKMPMM{d}{\varepsilon}{k}{t}}{k^{2}}\right\} ,\label{eq:hel-t-lightc-res-1}\\
\mathcal{M}_{LIGHT}^{C}(-) & = & \,\,\,\,2\omega_{+}^{t}\left\{ -\frac{\BKM{d}{t}\BSSKP{b}{p}{\varepsilon}{\bar{u}}}{p^{2}}+\frac{\BKP{b}{\bar{u}}\BSSKM{d}{\varepsilon}{k}{t}}{k^{2}}\right\} ,\label{eq:hel-t-lightc-res-2}\end{eqnarray}
with $p=p_{g}-p_{\bar{u}}$ and $q=p_{g}-p_{d}$. 

Again in all above equations, we suppressed the common factor $\sqrt{2E_{u}}\sqrt{2E_{d}}\sqrt{2E_{b}}$,
the coupling constants ${\displaystyle g_{s}\left(\frac{g}{\sqrt{2}}\right)^{2}}$,
and the $W$ boson propagator ${\displaystyle \frac{1}{p_{W}^{2}-m_{W}^{2}+im_{W}\Gamma_{W}}}$
with $p_{W}=p_{t}-p_{b}$.

\subsubsection{NLO corrections to HEAVY}

The Feynman diagrams, that generate real emission contributions through
coupling a gluon to the heavy quark lines, are shown in Figs.~\ref{fig:real_t-chan}(g)
to (n). We separate the NLO HEAVY real emission corrections into the
following four categories: \begin{eqnarray*}
{\rm HEAVY-A} & : & bq\rightarrow q'gt,\qquad{\rm including\,\,(g)\,\,\,\, and\,\,(h),}\\
{\rm HEAVY-B} & : & b\bar{q}'\rightarrow\bar{q}gt,\qquad{\rm including\,\,(i)\,\,\,\,\, and\,\,(j),}\\
{\rm HEAVY-C} & : & qg\rightarrow q'\bar{b}t,\qquad{\rm including\,\,(k)\,\,\,\, and\,\,(l),}\\
{\rm HEAVY-D} & : & \bar{q}'g\rightarrow\bar{q}\bar{b}t,\qquad{\rm including\,\,(m)\,\, and\,\,(n).}\end{eqnarray*}
Denoting the helicity amplitude as $\mathcal{M}_{HEAVY}^{A,B,C,D}(\lambda_{t})$,
then the helicity amplitudes for HEAVY-A are:\begin{eqnarray}
\mathcal{M}_{HEAVY}^{A}(+) & = & 2\omega_{-}^{t}\frac{\BKP{u}{b}\BSSKPPMP{t}{\varepsilon^{*}}{q}{d}}{q^{2}-m_{t}^{2}}+2\, m_{t}\,\omega_{+}^{t}\frac{\BKP{u}{b}\BSKP{t}{\varepsilon^{*}}{d}}{q^{2}-m_{t}^{2}}\nonumber \\
 & + & 2\omega_{-}^{t}\frac{\BKPP{t}{d}\BSSKP{u}{p}{\varepsilon^{*}}{b}}{p^{2}},\label{eq:hel-t-heavya-res-1}\\
\mathcal{M}_{HEAVY}^{A}(-) & = & 2\omega_{+}^{t}\frac{\BKP{u}{b}\BSSKM{t}{\varepsilon^{*}}{q}{d}}{q^{2}-m_{t}^{2}}+2\, m_{t}\,\omega_{-}^{t}\frac{\BKP{u}{b}\BSKMMP{t}{\varepsilon^{*}}{d}}{q^{2}-m_{t}^{2}}\nonumber \\
 & + & 2\omega_{+}^{t}\frac{\BKM{t}{d}\BSSKP{u}{p}{\varepsilon^{*}}{b}}{p^{2}},\label{eq:hel-t-heavya-res-2}\end{eqnarray}
with $p=p_{b}-p_{g}$ and $q=p_{g}+p_{t}$. 

The helicity amplitudes for HEAVY-B are:\begin{eqnarray}
\mathcal{M}_{HEAVY}^{B}(+) & = & 2\omega_{-}^{t}\frac{\BKP{\bar{u}}{b}\BSSKPPMP{t}{\varepsilon^{*}}{q}{\bar{d}}}{q^{2}-m_{t}^{2}}+2\, m_{t}\,\omega_{+}^{t}\frac{\BKP{\bar{u}}{b}\BSKP{t}{\varepsilon^{*}}{\bar{d}}}{q^{2}-m_{t}^{2}}\nonumber \\
 & + & 2\omega_{-}^{t}\frac{\BKPP{t}{\bar{d}}\BSSKP{\bar{u}}{p}{\varepsilon^{*}}{b}}{p^{2}},\label{eq:hel-t-heavyb-res-1}\\
\mathcal{M}_{HEAVY}^{B}(-) & = & 2\omega_{+}^{t}\frac{\BKP{\bar{u}}{b}\BSSKM{t}{\varepsilon^{*}}{q}{\bar{d}}}{q^{2}-m_{t}^{2}}+2\, m_{t}\,\omega_{-}^{t}\frac{\BKP{\bar{u}}{b}\BSKMMP{t}{\varepsilon^{*}}{\bar{d}}}{q^{2}-m_{t}^{2}}\nonumber \\
 & + & 2\omega_{+}^{t}\frac{\BKM{t}{\bar{d}}\BSSKP{\bar{u}}{p}{\varepsilon^{*}}{b}}{p^{2}},\label{eq:hel-t-heavyb-res-2}\end{eqnarray}

The helicity amplitudes for HEAVY-C are:\begin{eqnarray}
\mathcal{M}_{HEAVY}^{C}(+) & = & 2\omega_{-}^{t}\frac{\BKP{u}{\bar{b}}\BSSKPPMP{t}{\varepsilon}{q}{d}}{q^{2}-m_{t}^{2}}-2\, m_{t}\,\omega_{+}^{t}\frac{\BKP{u}{\bar{b}}\BSKP{t}{\varepsilon}{d}}{q^{2}-m_{t}^{2}}\nonumber \\
 & - & 2\omega_{-}^{t}\frac{\BKPP{t}{d}\BSSKP{u}{p}{\varepsilon}{\bar{b}}}{p^{2}},\label{eq:hel-t-heavyc-res-1}\\
\mathcal{M}_{HEAVY}^{C}(-) & = & 2\omega_{+}^{t}\frac{\BKP{u}{\bar{b}}\BSSKM{t}{\varepsilon}{q}{d}}{q^{2}-m_{t}^{2}}-2\, m_{t}\,\omega_{-}^{t}\frac{\BKP{u}{\bar{b}}\BSKMMP{t}{\varepsilon}{d}}{q^{2}-m_{t}^{2}}\nonumber \\
 & - & 2\omega_{+}^{t}\frac{\BKM{t}{d}\BSSKP{u}{p}{\varepsilon}{\bar{b}}}{p^{2}},\label{eq:hel-t-heavyc-res-2}\end{eqnarray}
where $p=p_{g}-p_{\bar{b}}$ and $q=p_{g}-p_{t}$. 

The helicity amplitudes for HEAVY-D are:\begin{eqnarray}
\mathcal{M}_{HEAVY}^{D}(+) & = & 2\omega_{-}^{t}\frac{\BKP{\bar{u}}{\bar{b}}\BSSKPPMP{t}{\varepsilon}{q}{\bar{d}}}{q^{2}-m_{t}^{2}}-2\, m_{t}\,\omega_{+}^{t}\frac{\BKP{\bar{u}}{\bar{b}}\BSKP{t}{\varepsilon}{\bar{d}}}{q^{2}-m_{t}^{2}}\nonumber \\
 & - & 2\omega_{-}^{t}\frac{\BKPP{t}{\bar{d}}\BSSKP{\bar{u}}{p}{\varepsilon}{\bar{b}}}{p^{2}},\label{eq:hel-t-heavyd-res-1}\\
\mathcal{M}_{HEAVY}^{D}(-) & = & 2\omega_{+}^{t}\frac{\BKP{\bar{u}}{\bar{b}}\BSSKM{t}{\varepsilon}{q}{\bar{d}}}{q^{2}-m_{t}^{2}}-2\, m_{t}\,\omega_{-}^{t}\frac{\BKP{\bar{u}}{\bar{b}}\BSKMMP{t}{\varepsilon}{\bar{d}}}{q^{2}-m_{t}^{2}}\nonumber \\
 & - & 2\omega_{+}^{t}\frac{\BKM{t}{\bar{d}}\BSSKP{\bar{u}}{p}{\varepsilon}{\bar{b}}}{p^{2}},\label{eq:hel-t-heavyd-res-2}\end{eqnarray}
where $p=p_{g}-p_{\bar{b}}$ and $q=p_{g}-p_{t}$. 

Again in all above equations, we suppressed the common factor $\sqrt{2E_{u}}\sqrt{2E_{d}}\sqrt{2E_{b}}$,
the coupling constants ${\displaystyle g_{s}\left(\frac{g}{\sqrt{2}}\right)^{2}}$,
and the $W$ boson propagator ${\displaystyle \frac{1}{p_{W}^{2}-m_{W}^{2}+im_{W}\Gamma_{W}}}$
with $p_{W}=p_{u}-p_{d}$.

\subsubsection{NLO corrections to top quark decay}

\begin{figure}
\includegraphics[%
  scale=0.5]{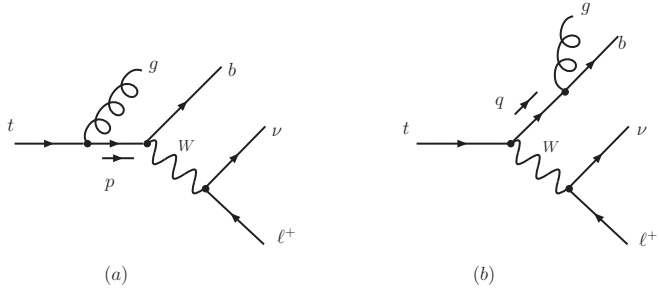}

\caption{Feynman diagrams of the real emission corrections to top quark decay
processes\label{fig: real_topdec}}
\end{figure}

The Feynman diagrams for NLO real emission corrections to the top
quark decay process are shown in Fig.~\ref{fig: real_topdec}. Denoting
the helicity amplitude as $\mathcal{M}_{DEC}(\lambda_{t})$, then\begin{eqnarray}
\mathcal{M}_{DEC}(+) & = & 2\,\omega_{-}^{t}\frac{\BKM{b}{\nu}\BSSKPMPP{e}{p}{\varepsilon^{*}}{t}}{p^{2}-m_{t}^{2}}+2\, m_{t}\,\omega_{+}^{t}\frac{\BKM{b}{\nu}\BSKP{e}{\varepsilon^{*}}{t}}{p^{2}-m_{t}^{2}}\nonumber \\
 & + & 2\,\omega_{-}^{t}\frac{\BKPP{e}{t}\BSSKM{b}{\varepsilon^{*}}{q}{\nu}}{q^{2}},\label{eq:hel-topdec-res-1}\\
\mathcal{M}_{DEC}(-) & = & 2\,\omega_{+}^{t}\frac{\BKM{b}{\nu}\BSSKP{e}{p}{\varepsilon^{*}}{t}}{p^{2}-m_{t}^{2}}+2\, m_{t}\,\omega_{-}^{t}\frac{\BKM{b}{\nu}\BSKPMM{e}{\varepsilon^{*}}{t}}{p^{2}-m_{t}^{2}}\nonumber \\
 & + & 2\,\omega_{+}^{t}\frac{\BKP{e}{t}\BSSKM{b}{\varepsilon^{*}}{q}{\nu}}{q^{2}},\label{eq:hel-topdec-res-2}\end{eqnarray}
with $p=p_{t}-p_{g}$ and $q=p_{b}+p_{g}$. We again suppressed the
common factor $\sqrt{2E_{e}}\sqrt{2E_{\nu}}\sqrt{2E_{b}}$, the coupling
constants ${\displaystyle g_{s}\left(\frac{g}{\sqrt{2}}\right)^{2}}$,
and the $W$ boson propagator ${\displaystyle \frac{1}{p_{W}^{2}-m_{W}^{2}+im_{W}\Gamma_{W}}}$
with $p_{W}=p_{e^{+}}+p_{\nu}$.

\section{NLO SCV form factors of the single top quark production and decay
processes}

In this section the analytical results of the effective form factors
are given in details together with the corresponding phase space boundary
conditions which slice the phase space of real emission corrections
into unresolved and resolved regions. Provided with such phase space
boundary conditions, one can use the helicity amplitudes given in
the previous section to perform numerical calculations. Since the
unresolved regions of massless partons differ from the ones of massive
partons, we separately consider  the massless and massive partons and present
the detailed derivations of the SCV form factors in Sec.~\ref{sub:SVC-init}
and Sec.~\ref{sub:SVC-FINAL}, respectively. For comparison, we present
our results in both the DREG and DRED schemes. We note that the form
factors and the crossing functions should be applied consistently
in any given scheme.

\subsection{NLO corrections to INIT\label{sub:SVC-init}}

Let us first examine the initial state corrections to the s-channel
single top quark process, cf. Fig.~\ref{fig:nlo}. After calculating
the effective matrix element with all the partons in the final state,
we cross the relevant partons into the initial state to obtain the
needed matrix element. In dimension $d=4-2\epsilon$, the NLO matrix
element for the vertex $q-\bar{q}'-W^{*}$ can be written as
\footnote{It yields $I_L=f_{1}^{q\bar{q}'\rightarrow W^{*}}$ in Eq.~(\ref{eq:form-s-init}).}
\begin{eqnarray}
M_{\mu}^{q\bar{q}'\rightarrow W^{*}} & = & \frac{ig}{\sqrt{2}}\bar{v}(\bar{q}^{\prime})[f_{1}^{q\bar{q}'\rightarrow W^{*}}\gamma_{\mu}]P_{L}u(q),\label{eq:Mat-ff-ini}
\end{eqnarray}
where $u(q)$($\bar{v}(\bar{q}^{\prime}))$ is the wave function
of $q(\bar{q}')$, $P_{L}=(1-\gamma_{5})/2$. 
The calculation of the virtual corrections for the vertex $q-\bar{q}^{\prime}-W^{*}$
is straightforward and after renormalization it yields
\begin{eqnarray}
f_{1}^{q\bar{q}'\rightarrow W({\rm virt})} & = &
\frac{\alpha_{s}}{4\pi}C_{F}C_{\epsilon}\left\{
-\frac{2}{\epsilon^{2}}+\frac{2}{\epsilon}\ln\frac{\hat{s}}{m_{t}^{2}}-\frac{3}{\epsilon}\right.\nonumber \\
 &\,  & \left.+\frac{4\pi^{2}}{3}+3\ln\frac{\hat{s}}{m_{t}^{2}}-\ln^{2}\frac{\hat{s}}{m_{t}^{2}}+I_{{\rm scheme}}^{q\bar{q}'\rightarrow W({\rm virt})}\right\} ,\label{eq:I-virt-init}
\end{eqnarray}
 where $\hat{s}=2p_{q}\cdot p_{\bar{q}^{\prime}}$, $C_{F}=4/3$,
${\displaystyle C_{\epsilon}=\left(\frac{4\pi\mu^{2}}{m_{t}^{2}}\right)^{\epsilon}\Gamma(1+\epsilon)}$,
and the scheme dependent term $I_{\,{\rm scheme}}^{\, q\bar{q}'\rightarrow W({\rm virt})}$
is \begin{equation}
I_{{\rm \, Scheme}}^{\, q\bar{q}'\rightarrow W({\rm virt})}=\begin{cases}
-8 & {\rm in~DREG~scheme},\\
-7 & {\rm in~DRED~scheme}.\end{cases}\end{equation}
We have neglected all the possible imaginary parts in the above result
and also in what follows, because they do not contribute to cross sections
up to the NLO. Note that
the tree level amplitude corresponds to setting $f_{1}^{q\bar{q}'\rightarrow W}=1$
in Eq.~(\ref{eq:Mat-ff-ini}). 

In the phase space slicing method, the soft and collinear singularities
in the virtual corrections, the poles of $\epsilon$ in Eq.~(\ref{eq:I-virt-init}),
should be canceled by the unresolved real emission corrections from
processes shown in Eqs.~(\ref{qqo})-(\ref{qqt}). Below, we will
partition the phase space of the real emission corrections to calculate
the unresolved contribution.

\begin{figure}
\includegraphics[%
  scale=0.6]{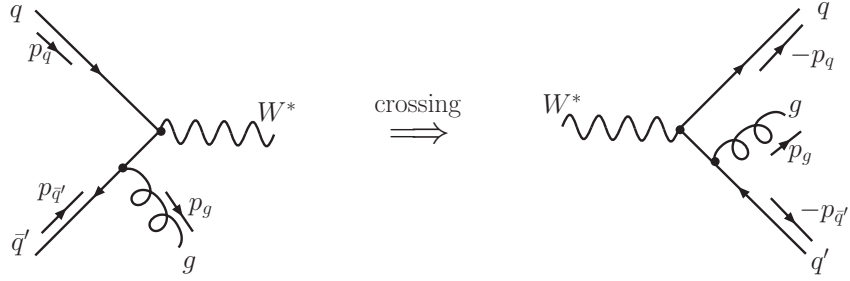}

\caption{Illustration for crossing the initial state partons into the final
state in the process $q\bar{q}^{\prime}\rightarrow W^{*}g$.\label{fig:pss_crossing}}
\end{figure}

\begin{figure}
\includegraphics[%
  scale=0.8]{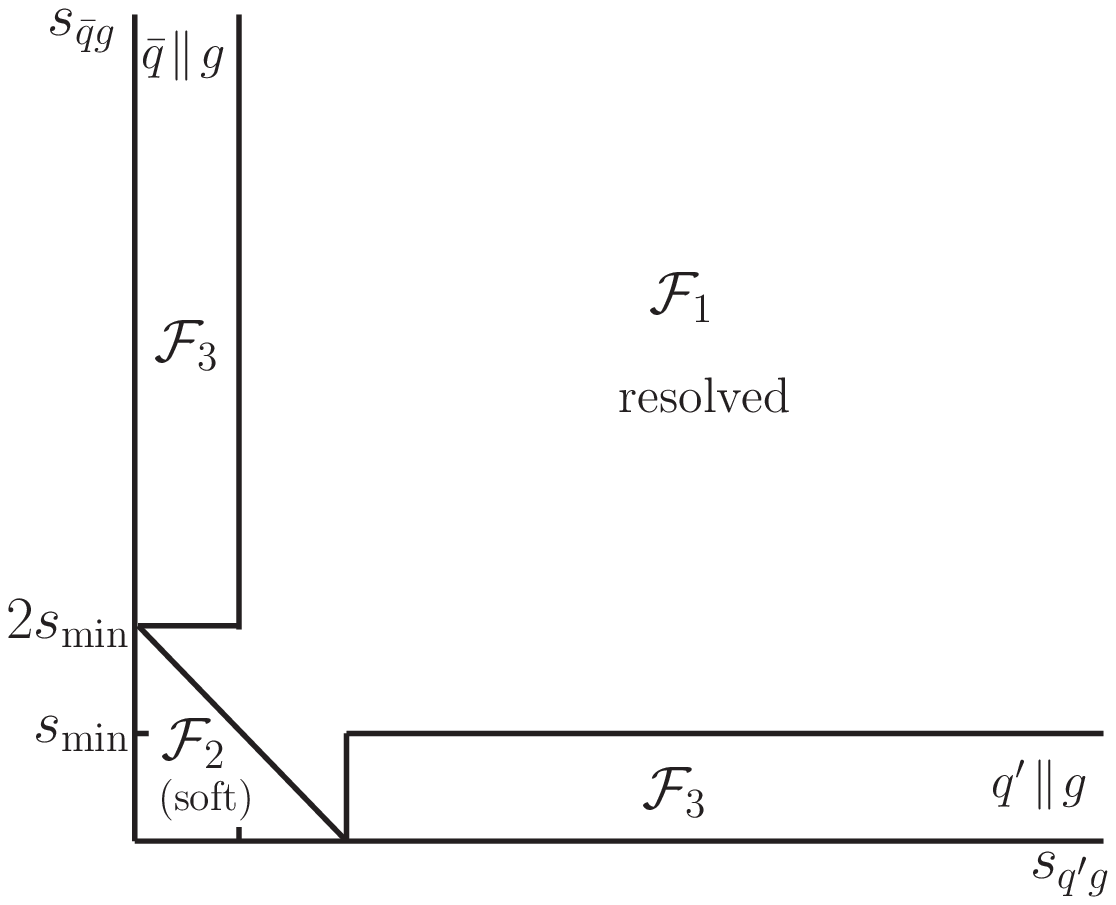}

\caption{The $s_{\bar{q}g}-s_{q^{\prime}g}$ plane for quark pair annihilation
to virtual $W$-boson showing the delineation into soft ($\mathcal{F}_{2}$),
 collinear ($\mathcal{F}_{2}$) and resolved region ($\ \mathcal{F}_{1}$).
\label{fig:phase_slicing_init}}
\end{figure}

As an example, let us examine the $q\bar{q}\rightarrow W^{*}g$ process.
After crossing all the initial state partons of the process $q\bar{q}^{\prime}\rightarrow W^{*}g$
into the final state, the particles' momenta are assigned as in Fig.~\ref{fig:pss_crossing}
which implies the crossed process $W^{*}\rightarrow\bar{q}q'g$.
Let us consider the whole phase space of the process $W^{*}\rightarrow\bar{q}q^{\prime}g$
as the identity and partition it into three regions as shown in Fig.~\ref{fig:phase_slicing_init}:\begin{eqnarray}
1 & \equiv & \Theta(\left|s_{\bar{q}g}\right|+\left|s_{q'g}\right|-2s_{min})+\Theta(2s_{min}-\left|s_{\bar{q}g}\right|-\left|s_{q'g}\right|)\nonumber \\
 & - & \Theta(\left|s_{\bar{q}g}\right|-2s_{min})\Theta(s_{min}-\left|s_{q'g}\right|)-\Theta(\left|s_{q'g}\right|-2s_{min})\Theta(s_{min}-\left|s_{\bar{q}g}\right|)\nonumber \\
 & + & \Theta(\left|s_{\bar{q}g}\right|-2s_{min})\Theta(s_{min}-\left|s_{q'g}\right|)+\Theta(\left|s_{q'g}\right|-2s_{min})\Theta(s_{min}-\left|s_{\bar{q}g}\right|)\nonumber \\
 & = & \mathcal{F}_{1}+\mathcal{F}_{2}+\mathcal{F}_{3},\label{eq:pss_condition_init}\end{eqnarray}
where\begin{eqnarray}
\mathcal{F}_{1} & = & \Theta(\left|s_{\bar{q}g}\right|+\left|s_{q'g}\right|-2s_{min})\nonumber \\
 &  & -\Theta(\left|s_{\bar{q}g}\right|-2s_{min})\Theta(s_{min}-\left|s_{q'g}\right|)-\Theta(\left|s_{q'g}\right|-2s_{min})\Theta(s_{min}-\left|s_{\bar{q}g}\right|),\label{eq:pss_condition_init-1}\\
\mathcal{F}_{2} & = & \Theta(2s_{min}-\left|s_{\bar{q}g}\right|-\left|s_{q'g}\right|),\label{eq:pss_condition_init-2}\\
\mathcal{F}_{3} & = & \Theta(\left|s_{\bar{q}g}\right|-2s_{min})\Theta(s_{min}-\left|s_{q'g}\right|)+\Theta(\left|s_{q'g}\right|-2s_{min})\Theta(s_{min}-\left|s_{\bar{q}g}\right|).\label{eq:pss_condition_init-3}\end{eqnarray}
Here $\Theta$ is the Heaviside step function and $s_{ij}=2p_{i}\cdot p_{j}$,
where $p_{i}$ is the four-momentum of the particle $i$. In the phase
space region constrained by function $\mathcal{F}_{1}$ (resolved),
there is no soft and collinear divergencies, therefore it can be calculated
in four dimensions numerically. The soft region is defined by the
function $\mathcal{F}_{2}$, which have both the soft and collinear
divergencies. The collinear regions is defined by the function $\mathcal{F}_{3}$
as shown in Fig.~\ref{fig:phase_slicing_init} which only have the
collinear singularities but no soft singularities. In function $\mathcal{F}_{3}$,
the first term denotes the collinear region of $g\parallel q'$ and
the second term represents the collinear region of $g\parallel\bar{q}$. 

Under the soft approximation, i.e. in the soft region ($\mathcal{F}_{2}$),
the squared matrix element can be written as a factor multiplying
the squared Born matrix element:\begin{equation}
\Theta(2s_{min}-\left|s_{\bar{q}g}\right|-\left|s_{q'g}\right|)\left|\mathcal{M}(W^{*}\rightarrow\bar{q}q^{\prime}g)\right|^{2}\xrightarrow{p_{g}\rightarrow0}\hat{f}_{s}^{W^{*}\rightarrow\bar{q}q^{\prime}g}\left|\mathcal{M}(W^{*}\rightarrow\bar{q}q^{\prime})\right|^2,\end{equation}
where we have defined the \emph{eikonal} factor $\hat{f}_{s}^{W^{*}\rightarrow\bar{q}q^{\prime}g}$
as \begin{equation}
\hat{f}_{s}^{W^{*}\rightarrow\bar{q}q^{\prime}g}=g_{s}C_{F}\mu^{2\epsilon}\frac{4(2p_{\bar{q}}\cdot p_{q^{\prime}})}{(2p_{\bar{q}}\cdot p_{g})(2p_{q^{\prime}}\cdot p_{g})}.\end{equation}
It is very simple to analytically integrate the eikonal factors $\hat{f}_{s}^{W^{*}\rightarrow\bar{q}q^{\prime}g}$
in $d$ dimensions over the soft gluon momentum~\cite{Brandenburg:1997pu} and
get the soft factor\begin{eqnarray}
I_{{\rm soft}}^{W^{*}\rightarrow\bar{q}q^{\prime}g} & = & \frac{g_{s}^{2}}{16\pi^{2}}\frac{C_{F}}{\Gamma(1-\epsilon)}\left(\frac{4\pi\mu_{d}^{2}}{s_{min}}\right)^{2}\left\{ \frac{2}{\epsilon^{2}}-\frac{4\ln2}{\epsilon}-\frac{2}{\epsilon}\ln\left(\frac{s_{min}}{\hat{s}}\right)\right.\nonumber \\
 &  & \left.+4\ln^{2}2-\frac{\pi^{2}}{3}+\ln^{2}\left(\frac{s_{min}}{\hat{s}}\right)+4\ln2\,\ln\left(\frac{s_{min}}{\hat{s}}\right)\right\} .\label{eq:I-soft-init}\end{eqnarray}

In addition to being singular in the soft gluon region, the matrix
elements are also singular in the collinear region ($\mathcal{F}_{3}$)
where the matrix elements exhibit an overall factorization. In the
limit $g\parallel\bar{q}$, we define 
\begin{equation}
p_{g}\xrightarrow{g\parallel\bar{q}}\xi p_{h},\qquad p_{\bar{q}}\xrightarrow{g\parallel\bar{q}}(1-\xi)p_{h},
\end{equation}
with $p_{h}=p_{g}+p_{\bar{q}}.$ In this limit, \begin{equation}
\Theta(\left|s_{q'g}\right|-2s_{min})\Theta(s_{min}-\left|s_{\bar{q}g}\right|)\left|\mathcal{M}(W^{*}\rightarrow\bar{q}q'g)\right|^{2}\xrightarrow{g\parallel\bar{q}}\hat{c}^{\bar{q}g\rightarrow\bar{q}}\left|\mathcal{M}(W^{*}\rightarrow\bar{q}q')\right|^{2},\end{equation}
where the collinear factor $\hat{c}^{\bar{q}g\rightarrow\bar{q}}$
is defined as~\cite{Brandenburg:1997pu}\begin{equation}
\hat{c}^{\bar{q}g\rightarrow\bar{q}}=g_{s}^{2}\mu^{2\epsilon}C_{F}\frac{P^{\bar{q}g\rightarrow\bar{q}}(\xi)}{2p_{g}\cdot p_{\bar{q}}}.\end{equation}
The function $P^{\bar{q}g\rightarrow\bar{q}}$ is related to the Altarelli-Parisi
splitting function, which depends on the regularization scheme. In this paper,
we adopt two schemes: the conventional dimensional regularization (DREG) scheme 
and the dimensional reduction (DRED) scheme. We have
\begin{equation}
P^{\bar{q}g\rightarrow\bar{q}}(\xi)=\begin{cases}
2\,{\displaystyle \frac{1+\xi^{2}-\epsilon(1-\xi)^{2}}{1-\xi},} & {\rm in~DREG~scheme},\\
2\,{\displaystyle \frac{1+\xi^{2}}{1-\xi},} & {\rm in~DRED~scheme}.\end{cases}\label{eq:Spliciting-kernel}\end{equation}
After integrating over the collinear phase space~\cite{Brandenburg:1997pu} for
the case $g\parallel q^{\prime}$, we obtain the collinear factor
\begin{eqnarray}
I_{{\rm col}}^{W^{*}\rightarrow\bar{q}q^{\prime}g} & = & \frac{g_{s}^{2}}{16\pi^{2}}\frac{C_{F}}{\Gamma(1-\epsilon)}\left(\frac{4\pi\mu_{d}^{2}}{s_{min}}\right)^{\epsilon}\nonumber \\
 & \times & \left\{ \frac{4}{\epsilon}\ln\left(\frac{2s_{min}}{\hat{s}}\right)+\frac{3}{\epsilon}-\frac{2\pi^{2}}{3}-2\ln^{2}\left(\frac{2s_{min}}{\hat{s}}\right)+I_{{\rm Scheme}}^{W^{*}\rightarrow\bar{q}q^{\prime}({\rm col})}\right\}, \label{eq:I-col-init}
 \end{eqnarray}
where the scheme dependent factor $I_{{\rm Scheme}}^{W^{*}\rightarrow\bar{q}q^{\prime}({\rm col})}$
is \begin{equation}
I_{{\rm Scheme}}^{W^{*}\rightarrow\bar{q}q^{\prime}({\rm col})}=\begin{cases}
7 & {\rm in~DREG~scheme},\\
6 & {\rm in~DRED~scheme}.\end{cases}\end{equation}

Summing over the soft and collinear factors and crossing the needed partons
into the initial state, we get the contributions to $\mathcal{M}^{q\bar{q}^{\prime}\rightarrow W^{*}}$
from the unresolved real (soft+collinear) corrections from processes
(\ref{qqo})-(\ref{qqt}) as following:
\begin{eqnarray}
f_{1}^{q\bar{q}'\rightarrow W({\rm real})} & = & \frac{\alpha_{s}}{4\pi}C_{F}C_{\epsilon}\left(\frac{2}{\epsilon^{2}}-\frac{2}{\epsilon}\ln\frac{\hat{s}}{m_{t}^{2}}+\frac{3}{\epsilon}-\frac{4\pi^{2}}{3}+2\ln^{2}2\right. \nonumber \\  
 &  & \left.+3\ln\frac{m_{t}^{2}}{s_{min}}-2\ln^{2}\frac{\hat{s}}{s_{min}}+\ln^{2}\frac{\hat{s}}{m_{t}^{2}}+I_{{\rm Scheme}}^{q\bar{q}'\rightarrow W({\rm real})}\right),
 \end{eqnarray}
where the scheme dependent term $I_{{\rm \, Scheme}}^{\, q\bar{q}'\rightarrow W({\rm real})}$
is 
\begin{equation}
I_{{\rm \, Scheme}}^{\, q\bar{q}'\rightarrow W({\rm real})}=\begin{cases}
7 & {\rm in~DREG~scheme},\\
6 & {\rm in~DRED~scheme}.\end{cases}
\end{equation}
It is clear that the divergencies of $f_{1}^{q\bar{q}'\rightarrow W({\rm virt})}$
and $f_{1}^{q\bar{q}'\rightarrow W({\rm real})}$ cancel with each
other and the sum is finite and $s_{min}$ dependent. The remaining
unresolved real corrections for $q\bar{q}^{\prime}\rightarrow W^{*}$
are included through the process independent, but $s_{min}$ and factorization
scheme dependent universal crossing functions.

The corrections from the resolved regions of processes (\ref{qqo})-(\ref{qqt})
can be obtained by multiplying the following phase space slicing functions
to the corresponding phase space elements and matrix element squares
in the cross section calculations: \begin{eqnarray}
 &  & \biggl[\Theta(\left|s_{gq}\right|+\left|s_{g\bar{q'}}\right|-2s_{min})-\Theta(\left|s_{gq}\right|-2s_{min})\Theta(s_{min}-\left|s_{g\bar{q'}}\right|) \nonumber \\
 &  & \,\,\,\,\,\,\,\,\,\,\,\,\,\,\,\,\,\,\,\,\,\,\,\,\,\,\,\,\,\,-\Theta(\left|s_{g\bar{q'}}\right|-2s_{min})\Theta(s_{min}-\left|s_{gq}\right|)\biggr]~~\,\,\,\,{\textrm{for }}~~q\bar{q}'\rightarrow W^{*}g\rightarrow t\bar{b}g,~~~~ \\
 &  & \biggl[1-\Theta(s_{min}-\left|s_{gq'}\right|)\biggr]\,\,\,\,\,\,\,\,\,\,\,\,\,\,\,\,\,\,\,\,\,\,\,\,\,\,\,\,\,\,\,\,\,\,\,\,\,\,\,\,\,\,\,\,\,\,\,\,\,\,\,\,\,\,\,\,\,\,\,\,\,\,\,\,\,\,\,~~{\textrm{for }}~~qg\rightarrow W^{*}q'\rightarrow t\bar{b}q',~~~~\\
 &  & \biggl[1-\Theta(s_{min}-\left|s_{g\bar{q}}\right|)\biggr]\,\,\,\,\,\,\,\,\,\,\,\,\,\,\,\,\,\,\,\,\,\,\,\,\,\,\,\,\,\,\,\,\,\,\,\,\,\,\,\,\,\,\,\,\,\,\,\,\,\,\:\,\,\,\,\,\,\,\,\,\,\,\,\,\,\,\,\,~~{\textrm{for }}~~g\bar{q}'\rightarrow W^{*}\bar{q}\rightarrow t\bar{b}\bar{q}.\end{eqnarray}
 The $\Theta$ functions ensure the amplitude squares to be finite
in four dimensions. Therefore, they can be calculated numerically.

\subsection{NLO corrections to FINAL\label{sub:SVC-FINAL}}

Now we examine the final state corrections to $W_{\mu}^{*}-t-\bar{b}$.
The NLO matrix element for the $W^{*}-t-b$ vertex can be written as
\footnote{It yields $F_1^{L*}=f_1^{W^{*}\rightarrow t\bar{b}}$, 
$F_2^{R*}=2f_2^{W^{*}\rightarrow t\bar{b}}/m_t$ and 
$F_1^{R*}=F_2^{L*}=0$ in Eq.~(\ref{eq:form-s-fnal}).}
\begin{eqnarray}
\mathcal{M}_{\mu}^{W^{*}\rightarrow t\bar{b}} & = & \frac{ig}{\sqrt{2}}\bar{u}(t)[f_{1}^{W^{*}\rightarrow t\bar{b}}\gamma_{\mu}-f_{2}^{W^{*}\rightarrow t\bar{b}}\frac{(p_{t}-p_{\bar{b}})_{\mu}}{m_{t}}]P_{L}v(\bar{b}).\end{eqnarray}
The above formula is valid only when $W$ boson is on-shell or off-shell
but coupled to massless quarks because we have neglected the term proportional
to $(p_{t}+p_{\bar{b}})_{\mu}$. At the tree level, $f_{1}^{W\rightarrow t\bar{b}}=1$
and $f_{2}^{W\rightarrow t\bar{b}}=0$. At the NLO, the virtual corrections
to $f_{1}^{W\rightarrow t\bar{b}}$ and $f_{2}^{W\rightarrow t\bar{b}}$
are, respectively,
\begin{eqnarray}
f_{1}^{W\rightarrow t\bar{b}({\rm virt})} & = & \frac{\alpha_{s}}{4\pi}C_{F}C_{\epsilon}\left\{ -\frac{1}{\epsilon^{2}}-\frac{5}{2\epsilon}+\frac{2}{\epsilon}\ln\frac{\hat{s}_{1}}{m_{t}^{2}}+\pi^{2}+2{\textrm{Li}}_{2}(\frac{\hat{s}}{\hat{s}_{1}})\right.\nonumber \\
 &  & \left.+3\ln\frac{\hat{s}_{1}}{m_{t}^{2}}-\frac{m_{t}^{2}}{\hat{s}}\ln\frac{\hat{s}_{1}}{m_{t}^{2}}-\ln^{2}\frac{\hat{s}_{1}}{m_{t}^{2}}+I_{\,{\rm Scheme}}^{\, W\rightarrow t\bar{b}({\rm virt})}\right\} ,\label{eq:I-virt-final-1}\\
f_{2}^{W\rightarrow t\bar{b}({\rm virt})} & = & \frac{\alpha_{s}}{4\pi}C_{F}C_{\epsilon}\left\{ \frac{m_{t}^{2}}{\hat{s}}\ln\frac{\hat{s}_{1}}{m_{t}^{2}}\right\} ,\label{eq:I-virt-final-2}
\end{eqnarray}
 where $\hat{s}=(p_{t}+p_{\bar{b}})^{2}$, $\hat{s}_{1}=2p_{t}\cdot p_{\bar{b}}=\hat{s}-m_{t}^{2}$
and the scheme dependent term $I_{\,{\rm Scheme}}^{\, W\rightarrow t\bar{b}({\rm virt})}$
is \begin{equation}
I_{{\rm \, Scheme}}^{\, W\rightarrow t\bar{b}({\rm virt})}=\begin{cases}
-6 & {\rm in~DREG~scheme,}\\
-5 & {\rm in~DRED~scheme.}\end{cases}\end{equation}

\begin{figure}
\includegraphics[%
  scale=0.8]{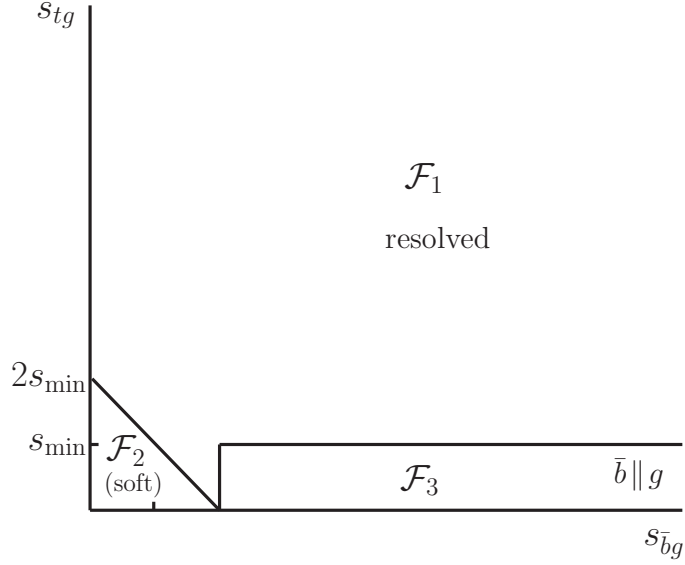}

\caption{The $s_{tg}-s_{\bar{b}g}$ plane for quark pair annihilation to virtual
$W$-boson showing the delineation into soft ($\mathcal{F}_{2}$), collinear ($\mathcal{F}_{2}$) and resolved region ($\mathcal{F}_{1}$).\label{fig:phase_slicing_final}}
\end{figure}

In the presence of massive top quark, the structure of collinear and
soft poles of FINAL matrix elements is completely different from the
massless case (INIT). The top quark mass serves as a regularizer for
collinear singularities. Thus, the matrix element contain fewer singular
structures. However, the presence of the top quark mass leads to more
complicated phase space integrals. Again, let us consider the whole
phase space of the process $W^{*}\rightarrow t\bar{b}g$ as the identity
and partition it into three regions as shown in Fig.~\ref{fig:phase_slicing_final}:\begin{eqnarray}
1 & \equiv & \Theta(s_{tg}+s_{\bar{b}g}-2s_{min})+\Theta(2s_{min}-s_{tg}-s_{\bar{b}g})\nonumber \\
 & + & \Theta(s_{tg}-2s_{min})\Theta(s_{min}-s_{\bar{b}g})-\Theta(s_{tg}-2s_{min})\Theta(s_{min}-s_{\bar{b}g})\nonumber \\
 & = & \mathcal{F}_{1}+\mathcal{F}_{2}+\mathcal{F}_{3},\label{eq:pss_condition_fnal}\end{eqnarray}
where\begin{eqnarray}
\mathcal{F}_{1} & = & \Theta(s_{tg}+s_{\bar{b}g}-2s_{min})-\Theta(s_{tg}-2s_{min})\Theta(s_{min}-s_{\bar{b}g}),\label{eq:pss_condition_fnal-1}\\
\mathcal{F}_{2} & = & \Theta(2s_{min}-s_{tg}-s_{\bar{b}g}),\label{eq:pss_condition_fnal-2}\\
\mathcal{F}_{3} & = & \Theta(s_{tg}-2s_{min})\Theta(s_{min}-s_{\bar{b}g}).\label{eq:pss_condition_fnal-3}\end{eqnarray}
Here again, we divide the phase space of process $W^{*}\rightarrow t\bar{b}g$
into three parts: the resolved region ($\mathcal{F}_{1}$), the soft
region ($\mathcal{F}_{2}$) and the collinear region ($\mathcal{F}_{3}$).
Moreover, the phase space boundary conditions are much simpler than
the case of massless partons, cf. Eqs.~(\ref{eq:pss_condition_init-1})-(\ref{eq:pss_condition_init-3}).

Under the soft approximation, in the soft region ($\mathcal{F}_{2}$),
the squared matrix element can be written as a factor multiplying
the squared Born matrix element:\begin{equation}
\Theta(2s_{min}-s_{tg}-s_{\bar{b}g})\left|\mathcal{M}(W^{*}\rightarrow t\bar{b}g)\right|^{2}\xrightarrow{p_{g}\rightarrow0}\hat{f}_{s}^{W^{*}\rightarrow tbg}\left|\mathcal{M}(W^{*}\rightarrow t\bar{b})\right|^2,
\end{equation}
where we have defined the \emph{eikonal} factor $\hat{f}_{s}^{W^{*}\rightarrow t\bar{b}g}$
as~\cite{Brandenburg:1997pu}:
\begin{equation}
\hat{f}_{s}^{W^{*}\rightarrow t\bar{b}g}=g_{s}C_{F}\mu^{2\epsilon}\left[\frac{4(2p_{t}\cdot p_{\bar{b}})}{(2p_{t}\cdot p_{g})(2p_{\bar{b}}\cdot p_{g})}-\frac{4m_{t}^{2}}{(2p_{t}\cdot p_{g})^{2}}\right].
\end{equation}
Integrating the eikonal factors $\hat{f}_{s}^{W^{*}\rightarrow t\bar{b}g}$
in $d$ dimensions over the soft gluon momentum~\cite{Brandenburg:1997pu}, we
get the soft factor\begin{eqnarray}
I_{{\rm soft}}^{W^{*}\rightarrow t\bar{b}g} & = & \frac{g_{s}^{2}}{16\pi^{2}}\frac{C_{F}}{\Gamma(1-\epsilon)}\left(\frac{4\pi\mu_{d}^{2}}{s_{min}}\right)^{2}\left(\frac{s_{min}}{\hat{s}_{1}+m_{t}^{2}}\right)^{-\epsilon}\nonumber \\
 & \times & \left\{ \frac{1}{\epsilon^{2}}--\frac{1}{\epsilon}\left[\ln\left(1+\frac{\hat{s}_{1}}{m_{t}^{2}}\right)+2\ln2-1\right]\right.\nonumber \\
 &  & -\frac{\pi^{2}}{6}+2\ln^{2}2-2\ln2+\left[2\ln2+\frac{\hat{s}_{1}+2m_{t}^{2}}{\hat{s}_{1}}\right]\ln\left(1+\frac{\hat{s}_{1}}{m_{t}^{2}}\right)\nonumber \\
 &  & \left.-\frac{1}{2}\ln^{2}\left(1+\frac{\hat{s}_{1}}{m_{t}^{2}}\right)-2{\rm Li_{2}\left(\frac{\hat{s}_{1}}{\hat{s}_{1}+m_{t}^{2}}\right)}\right\} .\label{eq:I-soft-final}\end{eqnarray}

In the collinear region ($\mathcal{F}_{3}$), where $g\parallel\bar{b}$,
the matrix elements exhibit an overall factorization as\begin{equation}
\Theta(s_{tg}-2s_{min})\Theta(s_{min}-s_{\bar{b}g})\left|\mathcal{M}(W^{*}\rightarrow t\bar{b}g)\right|^{2}\xrightarrow{g\parallel\bar{b}}\hat{c}^{\bar{b}g\rightarrow\bar{b}}(\xi)\left|\mathcal{M}(W^{*}\rightarrow t\bar{b})\right|^{2},\end{equation}
where the collinear factor $\hat{c}^{\bar{b}g\rightarrow\bar{b}}$
is defined as~\cite{Brandenburg:1997pu}:
\begin{equation}
\hat{c}^{\bar{b}g\rightarrow\bar{b}}=g_{s}^{2}\mu^{2\epsilon}C_{F}\left[\frac{P^{\bar{b}g\rightarrow\bar{b}}(\xi)}{2p_{\bar{b}}\cdot p_{g}}-\frac{4m_{t}^{2}}{\left(2p_{t}\cdot p_{g}\right)^{2}}\right],
\end{equation}
in which $P^{\bar{b}g\rightarrow\bar{b}}(\xi)$ is same as Eq.~(\ref{eq:Spliciting-kernel}).
Integrating over the collinear phase space~\cite{Brandenburg:1997pu}, we get
the collinear factor\begin{eqnarray}
I_{{\rm col}}^{W^{*}\rightarrow t\bar{b}g} & = & \frac{g_{s}^{2}}{16\pi^{2}}\frac{C_{F}}{\Gamma(1-\epsilon)}\left(\frac{4\pi\mu_{d}^{2}}{s_{min}}\right)^{\epsilon}\nonumber \\
 & \times & \left\{ \frac{2}{\epsilon}\left[\ln^{2}\left(\frac{2s_{min}}{\hat{s}}\right)+\frac{3}{4}\right]-\frac{\pi^{2}}{3}-\ln^{2}\left(\frac{2s_{min}}{\hat{s}}\right)-\frac{m_{t}^{2}}{\hat{s}_{1}}+I_{{\rm Scheme}}^{W^{*}\rightarrow t\bar{b}g({\rm col})}\right\}, \label{eq:I-col-final}\end{eqnarray}
where the scheme dependent factor $I_{{\rm Scheme}}^{W^{*}\rightarrow t\bar{b}g({\rm col})}$
is 
\begin{equation}
I_{{\rm Scheme}}^{W^{*}\rightarrow t\bar{b}g({\rm col})}=\begin{cases}
{\displaystyle \frac{7}{2}} & {\rm in~DREG~scheme},\\
3 & {\rm in~DRED~scheme}.\end{cases}
\end{equation}

Summing over the soft and collinear factor, we get the contributions to
$\mathcal{M}^{W^{*}\rightarrow t\bar{b}}$ from the unresolved real
(soft+collinear) corrections as following:
\begin{eqnarray}
f_{1}^{W^{*}\rightarrow t\bar{b}g({\rm real})} & = & \frac{\alpha_{s}}{4\pi}C_{F}C_{\epsilon}\left\{ \frac{1}{\epsilon^{2}}+\frac{5}{2\epsilon}-\frac{2}{\epsilon}\ln\frac{\hat{s}_{1}}{m_{t}^{2}}-\frac{2\pi^{2}}{3}+\ln^{2}2-2\ln2\right.\nonumber\\ 
 &  & -2{\textrm{Li}}_{2}(\frac{\hat{s}_{1}}{\hat{s}})-\frac{7}{2}\ln\frac{s_{min}}{m_{t}^{2}}+2\ln2\ln\frac{\hat{s}_{1}}{m_{t}^{2}}+(2+\frac{2m_{t}^{2}}{\hat{s}_{1}})\ln\frac{\hat{s}}{m_{t}^{2}}\nonumber\\ 
 &  & \left.-\ln^{2}\frac{\hat{s}}{m_{t}^{2}}-\ln^{2}\frac{\hat{s}_{1}}{s_{min}}+2\ln\frac{\hat{s}_{1}}{m_{t}^{2}}\ln\frac{s_{min}}{m_{t}^{2}}-\frac{m_{t}^{2}}{\hat{s}_{1}}+I_{\,{\rm Scheme}}^{\, W^{*}\rightarrow t\bar{b}g({\rm real})}\right\} ,\end{eqnarray}
where the scheme dependent term $I_{{\rm \, Scheme}}^{\, W^{*}\rightarrow t\bar{b}g({\rm real})}$
is\begin{equation}
I_{\,{\rm Scheme}}^{\, W^{*}\rightarrow t\bar{b}g({\rm real})}=\begin{cases}
{\displaystyle \frac{7}{2}} & {\rm in~DREG~scheme},\\
3 & {\rm in~DRED~scheme}.\end{cases}\end{equation}
 The correction from the resolved regions of process (\ref{tb}) can
be obtained by multiplying the following phase space slicing functions
to the corresponding phase space elements and matrix element squares
in the cross section calculations: \begin{equation}
\biggl[\Theta(s_{tg}+s_{\bar{b}g}-2s_{min})-\Theta(s_{tg}-2s_{min})\Theta(s_{min}-s_{\bar{b}g})\biggr].\end{equation}

\subsection{NLO corrections to LIGHT}

The NLO matrix element for the $q-W^{*}_{\mu}-q'$ vertex can be written as
\footnote{It yields $I_L=f_{1}^{q\rightarrow W^{*}q'}$ in Eqs.~(\ref{eq:hel-t-light-scv-1})-(\ref{eq:hel-t-light-scv-2}).}
\begin{eqnarray}
M_{\mu}^{q\rightarrow W^{*}q'} & = & \frac{ig}{\sqrt{2}}\bar{u}(q^{\prime})[f_{1}^{q\rightarrow W^{*}q'}\gamma_{\mu}P_{L}]u(q),\end{eqnarray}
 where $u(q)$($\bar{u}(q^{\prime}))$ is the wave function of $q(q')$.
The virtual correction to $f_{1}^{q\rightarrow W^{*}q'}$ is \begin{eqnarray}
f_{1}^{q\rightarrow W^{*}q'({\rm virt})} & = & \frac{\alpha_{s}}{4\pi}C_{F}C_{\epsilon}\left\{ -\frac{2}{\epsilon^{2}}+\frac{2}{\epsilon}\ln(\frac{-\hat{t}}{m_{t}^{2}})-\frac{3}{\epsilon}\right.\nonumber \\
 &  & \left.\,\,\,\,\,\,\,\,\,\,\,\,\,\,\,\,\,\,\,\,\,\,\,\,+\frac{\pi^{2}}{3}+3\ln(\frac{-\hat{t}}{m_{t}^{2}})-\ln^{2}(\frac{-\hat{t}}{m_{t}^{2}})+I_{\,{\rm scheme}}^{\, q\rightarrow W^{*}q^{\prime}({\rm virt})}\right\} ,\label{eq:I-virt-light}\end{eqnarray}
 where $\hat{t}=-2p_{q}\cdot p_{q^{\prime}}$ and the scheme dependent
term $I_{\,{\rm Scheme}}^{\, q\rightarrow W^{*}q^{\prime}({\rm virt})}$
is \begin{equation}
I_{\,{\rm Scheme}}^{\, q\rightarrow W^{*}q^{\prime}({\rm virt})}=\begin{cases}
-8 & {\rm in~DREG~scheme},\\
-7 & {\rm in~DRED~scheme}.\end{cases}\end{equation}
The tree level amplitude can be obtained by setting $f_{1}^{q\rightarrow W^{*}q'}=1$.

We now consider the unresolved real correction to $f_{1}^{q\rightarrow W^{*}q'}$.
There are two processes that contribute to $q-W^{*}_{\mu}-q'$ vertex:

\begin{itemize}
\item $bq\rightarrow tq'g\,(I)$, in which the gluon only connects with
the light quark ($q$,$q^{\prime}$) line, cf. Eqs.~(\ref{qbtw})
and (\ref{qbarbtw}), 
\item $bg\rightarrow t\bar{q}q'$, cf. Eq.~(\ref{bg}). 
\end{itemize}
The soft and collinear divergent regions of $bq\rightarrow tq'g\,(I)$
can be constrained by the function\begin{eqnarray}
 &  & \biggl[\Theta(2s_{min}-\left|s_{qg}\right|-\left|s_{q'g}\right|)\nonumber\\ 
 &  & +\Theta(\left|s_{qg}\right|-2s_{min})\Theta(s_{min}-\left|s_{q'g}\right|)\nonumber\\ 
 &  & +\Theta(\left|s_{q'g}\right|-2s_{min})\Theta(s_{min}-\left|s_{qg}\right|)\biggl].\end{eqnarray}
 In the above function, the first term constrains $p_{g}$ to be soft
and the second and third terms restrict $p_{g}$ to be collinear with
$p_{q'}$ and $p_{q}$, respectively. The process $bg\rightarrow t\bar{q}q'$
has only collinear divergent phase space region which is projected
by \begin{equation}
\biggl[\Theta(s_{min}-\left|s_{\bar{q}g}\right|)+\Theta(s_{min}-\left|s_{q'g}\right|)\biggr],\end{equation}
 in which the two terms require $p_{g}$ to be collinear with $p_{q}$
and $p_{q'}$, respectively. After performing all the above constrained
phase space integrations analytically, one can get the contribution
to $f_{1}^{q\rightarrow W^{*}q'}$ from the unresolved real emission corrections
as:
\begin{eqnarray}
f_{1}^{q\rightarrow W^{*}q'({\rm real})} & = & \frac{\alpha_{s}}{4\pi}C_{F}C_{\epsilon}\left\{
\frac{2}{\epsilon^{2}}-\frac{2}{\epsilon}\ln(\frac{-\hat{t}}{m_{t}^{2}})+\frac{3}{\epsilon}-\frac{4\pi^{2}}{3}+2\ln^{2}2\right. \nonumber\\
 &  & \left.+3\ln\frac{m_{t}^{2}}{s_{min}}-2\ln^{2}(\frac{-\hat{t}}{s_{min}})+\ln^{2}(\frac{-\hat{t}}{m_{t}^{2}})+I_{\,{\rm Scheme}}^{\, q\rightarrow W^{*}q'({\rm real})}\right\} ,
 \end{eqnarray}
where the scheme dependent term $I_{{\rm \, Scheme}}^{\, q\rightarrow W^{*}q'({\rm real})}$
is\begin{equation}
I_{\,{\rm Scheme}}^{\, q\rightarrow W^{*}q^{\prime}({\rm real})}=\begin{cases}
7 & {\rm in~DREG~scheme},\\
6 & {\rm in~DRED~scheme}.\end{cases}\end{equation}
It is clear that the divergencies of $f_{1}^{q\rightarrow W^{*}q'({\rm virt})}$
and $f_{1}^{q\rightarrow W^{*}q'({\rm real})}$ cancel with each other
and the sum is finite and $s_{min}$ dependent. The remaining unresolved
real corrections for $q\rightarrow W^{*}q^{\prime}$ are included through
the process independent, but $s_{min}$ and factorization scheme dependent
universal crossing functions.

The resolved phase spaces without divergent regions are obtained by
multiplying the following function to the phase space:\begin{eqnarray}
 &  & \biggl[\Theta(\left|s_{gq}\right|+\left|s_{q^{\prime}g}\right|-2s_{min})-\Theta(\left|s_{gq}\right|-2s_{min})\Theta(s_{min}-\left|s_{q^{\prime}g}\right|)\nonumber\\ 
 &  & \,\,\,\,\,\,\,\,\,\,\,\,\,\,\,\,\,\,\,\,\,\,\,\,\,\,\,\,\,\,-\Theta(\left|s_{q^{\prime}g}\right|-2s_{min})\Theta(s_{min}-\left|s_{gq}\right|)\biggr]\,\,\,\,\,\,\,\,\,\,\,\,\,\,\,\,\,\,\,\,{\rm for}\,\,\,\, bq\rightarrow tq'g(I),\\
 &  & \biggl[1-\Theta(s_{min}-\left|s_{q'g}\right|)-\Theta(s_{min}-\left|s_{\bar{q}g}\right|)\biggr]\,\,\,\,\,\,\,\,\,\,\,\,\,\,\,\,\,\,\,\,\,\,\,\,\,\,\,\,\,\,\,\,\,\,\,\,{\rm for}\,\,\,\, bg\rightarrow t\bar{q}q'.\end{eqnarray}

\subsection{NLO corrections to HEAVY}

The NLO matrix element for the $b-W^{*}_{\mu}-t$ vertex can be written as
\footnote{It yields $H_1^{L*}=f_{1}^{bW^{*}\rightarrow t}$, 
$H_2^{R*}=2 f_{2}^{bW^{*}\rightarrow t}/m_t$ and 
$H_1^{R*}=H_2^{L*}=0$ in Eqs.~(\ref{eq:hel-t-heavy-scv-1})-(\ref{eq:hel-t-heavy-scv-4}).}
\begin{eqnarray}
M_{\mu}^{bW^{*}\rightarrow t} & = & \frac{ig}{\sqrt{2}}\bar{u}(t)[f_{1}^{bW^{*}\rightarrow t}\gamma_{\mu}-f_{2}^{bW^{*}\rightarrow t}\frac{(p_{b}+p_{t})_{\mu}}{m_{t}}]P_{L}u(b).\end{eqnarray}
The above formula is valid only when $W$ boson is on-shell or off-shell
but coupled to massless quarks because we have neglected the term proportional
to $(p_{t}-p_{b})_{\mu}$. At the LO, $f_{1}^{bW^{*}\rightarrow t}=1$ and 
$f_{2}^{bW^{*}\rightarrow t}=0$. At the NLO, the virtual correction to
$f_{1}^{bW^{*}\rightarrow t}$ and $f_{2}^{bW^{*}\rightarrow t}$ are
\begin{eqnarray}
f_{1}^{bW^{*}\rightarrow t({\rm virt})} & = & \frac{\alpha_{s}}{4\pi}C_{F}C_{\epsilon}\left\{ -\frac{1}{\epsilon^{2}}-\frac{5}{2\epsilon}+\frac{2}{\epsilon}\ln(\frac{-\hat{t}_{1}}{m_{t}^{2}})+2{\textrm{Li}}_{2}(\frac{\hat{t}}{\hat{t}_{1}})\right.\nonumber\\ 
 &  & \left.+3\ln(\frac{-\hat{t}_{1}}{m_{t}^{2}})-\frac{m_{t}^{2}}{\hat{t}}\ln(\frac{-\hat{t}_{1}}{m_{t}^{2}})-\ln^{2}(\frac{-\hat{t}_{1}}{m_{t}^{2}})+I_{\,{\rm Scheme}}^{\, bW^{*}\rightarrow t({\rm virt})}\right\} ,\\
f_{2}^{bW^{*}\rightarrow t({\rm virt})} & = & \frac{\alpha_{s}}{4\pi}C_{F}C_{\epsilon}\left\{ \frac{m_{t}^{2}}{\hat{t}}\ln(\frac{-\hat{t}_{1}}{m_{t}^{2}})\right\} ,
\end{eqnarray}
where $\hat{t}_{1}=\hat{t}-m_{t}^{2}=-2p_{b}\cdot p_{t}$, and the
scheme dependent term $I_{\,{\rm Scheme}}^{\, bW^{*}\rightarrow t({\rm virt})}$
is \begin{equation}
I_{\,{\rm Scheme}}^{\, bW^{*}\rightarrow t({\rm virt})}=\begin{cases}
-6 & {\rm in~DREG~scheme},\\
-5 & {\rm in~DRED~scheme}.\end{cases}\end{equation}

We now consider the unresolved real correction to $f_{1}^{bW^{*}\rightarrow t}$.
The unresolved real correction to $f_{1}^{bW^{*}\rightarrow t}$ comes
from the soft and collinear regions of the following three processes:

\begin{itemize}
\item $bq\rightarrow tq'g\,(II)$, in which the gluon only connect with
the heavy quark ($t$,$b$) line, cf. Eqs.~(\ref{qbtw}) and (\ref{qbarbtw}), 
\item $qg\rightarrow tq'\bar{b}$, cf. Eq.~(\ref{qg}),
\item $\bar{q}'g\rightarrow\bar{q}\bar{b}t$, cf. Eq.~(\ref{qbarg}).
\end{itemize}
The soft and collinear divergent regions of $bq\rightarrow tq'g\,(II)$
are sliced out by
\begin{equation}
\biggl[\Theta(2s_{min}-\left|s_{bg}\right|-\left|s_{tg}\right|)+\Theta(\left|s_{tg}\right|-2s_{min})\Theta(s_{min}-\left|s_{bg}\right|)\biggr].
\end{equation}
The $qg\rightarrow tq'\bar{b}$ and $\bar{q}'g\rightarrow\bar{q}\bar{b}t$
processes both have the collinear divergent region restricted by
$\Theta(s_{min}-\left|s_{\bar{b}g}\right|)$. After integrating out
the soft and collinear regions, we get the contribution to the form
factor $f_{1}^{bW^{*}\rightarrow t}$ from the unresolved real emission
corrections as:
\begin{eqnarray}
f_{1}^{bW^{*}\rightarrow t({\rm real})} & = & \frac{\alpha_{s}}{4\pi}C_{F}C_{\epsilon}\left\{ \frac{1}{\epsilon^{2}}+\frac{5}{2\epsilon}-\frac{2}{\epsilon}\ln\frac{-\hat{t}_{1}}{m_{t}^{2}}-\frac{2\pi^{2}}{3}+\ln^{2}2-2\ln2-2{\textrm{Li}}_{2}(\frac{-\hat{t}_{1}}{m_{t}^{2}-\hat{t}_{1}})\right.\nonumber\\ 
 &  & -\frac{7}{2}\ln\frac{s_{min}}{m_{t}^{2}}+2\ln2\ln(\frac{-\hat{t}_{1}}{m_{t}^{2}})+(2-\frac{2m_{t}^{2}}{\hat{t}_{1}})\ln(1-\frac{\hat{t}_{1}}{m_{t}^{2}})-\ln^{2}(1-\frac{\hat{t}_{1}}{m_{t}^{2}}) \nonumber\\ 
 &  & \left.-\ln^{2}(\frac{-\hat{t}_{1}}{m_{t}^{2}})-\ln^{2}\frac{s_{min}}{m_{t}^{2}}+4\ln(\frac{-\hat{t}_{1}}{m_{t}^{2}})\ln\frac{s_{min}}{m_{t}^{2}}+\frac{m_{t}^{2}}{\hat{t}_{1}}+I_{\,{\rm Scheme}}^{\, bW^{*}\rightarrow t({\rm real})}\right\} ,
\end{eqnarray}
 where the scheme dependent term $I_{\,{\rm Scheme}}^{\, bW^{*}\rightarrow t({\rm real})}$
is \begin{equation}
I_{\,{\rm Scheme}}^{\, bW^{*}\rightarrow t({\rm real})}=\begin{cases}
{\displaystyle \frac{7}{2}} & {\rm in~DREG~scheme},\\
3 & {\rm in~DRED~scheme}.\end{cases}\end{equation}
and the remaining unresolved real corrections for $bW^{*}\rightarrow t$
are included through the process independent, but $s_{min}$ and factorization
scheme dependent universal crossing functions.

The resolved phase spaces without divergent regions are obtained by
multiplying the following function to the phase space:
\begin{eqnarray}
&\biggl[\Theta(\left|s_{bg}\right|+\left|s_{tg}\right|-2s_{min})-\Theta(\left|s_{tg}\right|-2s_{min})\Theta(s_{min}-\left|s_{bg}\right|)\biggr]  \,\,\,\, & {\rm for}\,\,\, bq\rightarrow tq'g\,(II),\\
&\biggl[1-\Theta(s_{min}-\left|s_{\bar{b}g}\right|)\biggr]   \,\,\,\, \,\,\,\, \,\,\,\, \,\,\,\,\,\,\,\, & {\rm for}\,\,\, qg\rightarrow tq'\bar{b}~\rm{and}\nonumber\\ 
&\,&~~~~~\bar{q}'g\rightarrow\bar{q}\bar{b}t.
\end{eqnarray}

\subsection{NLO corrections to the decay process $t\rightarrow Wb'$}

The NLO matrix element for the $t-W-b^{\prime}$ vertex can be written as
\footnote{It yields $D_1^{L}=f_{1}^{t\rightarrow Wb'}$, 
$D_2^{R}=-2 f_{2}^{t\rightarrow Wb'}/m_t$ and
$D_1^{R}=D_2^{L}=0$ in Eq.~(\ref{eq:form-topdec}).}
\begin{eqnarray}
M_{\mu}^{t\rightarrow Wb} & = & \frac{ig}{\sqrt{2}}\bar{u}(b')[f_{1}^{t\rightarrow Wb'}\gamma_{\mu}P_{L}+f_{2}^{t\rightarrow Wb'}\frac{(p_{t}+p_{b^{\prime}})_{\mu}}{m_{t}}P_{R}]u(t),\label{decay1}
\end{eqnarray}
where $P_{R}=(1+\gamma_{5})/2$ and $b^{\prime}$ denotes the bottom
quark from the top decay. At the tree level, $f_{1}^{t\rightarrow Wb'}=1$,
$f_{2}^{t\rightarrow Wb'}=0$. At the NLO, the virtual correction
to $f_{1}^{t\rightarrow Wb'}$ and $f_{2}^{t\rightarrow Wb'}$ are, respectively,
\begin{eqnarray}
f_{1}^{t\rightarrow Wb'({\rm virt})} & = & \frac{\alpha_{s}}{4\pi}C_{F}C_{\epsilon}\left\{ -\frac{1}{\epsilon^{2}}-\frac{5}{2\epsilon}+\frac{2}{\epsilon}\ln(1-\beta_{W})+2{\textrm{Li}}_{2}(\frac{\beta_{W}}{\beta_{W}-1})\right.\nonumber \\
 &  & \left.+\frac{3\beta_{W}-1}{\beta_{W}}\ln(1-\beta_{W})-\ln^{2}(1-\beta_{W})+I_{\,{\rm Scheme}}^{\, t\rightarrow Wb'({\rm virt})}\right\} ,\label{eq:I-virt-tdec-1}\\
f_{2}^{t\rightarrow Wb'({\rm virt})} & = & \frac{\alpha_{s}}{4\pi}C_{F}C_{\epsilon}\left\{ \frac{1}{\beta_{W}}\ln(1-\beta_{W})\right\} ,\label{eq:I-virt-tdec-2}\end{eqnarray}
where $\beta_{W}=m_{W}^{2}/m_{t}^{2}$, and the scheme dependent term
$I_{\,{\rm Scheme}}^{\, t\rightarrow Wb'({\rm virt})}$ is 
\begin{equation}
I_{\,{\rm Scheme}}^{\, t\rightarrow Wb'({\rm virt})}=\begin{cases}
-6 & {\rm in~DREG~scheme},\\
-5 & {\rm in~DRED~scheme}.\end{cases}
\end{equation}

The unresolved real correction to $f_{1}^{t\rightarrow Wb'}$ is obtained
by integrating out the soft and collinear regions of $t\rightarrow Wb'g$
which are sliced by \begin{equation}
\biggl[\Theta(2s_{min}-\left|s_{tg}\right|-\left|s_{b'g}\right|)+\Theta(\left|s_{tg}\right|-2s_{min})\Theta(s_{min}-\left|s_{b'g}\right|)\biggr].\label{eq:pss_condition_tdec-1}\end{equation}
 After integrating over the sliced regions, we get the
contribution to $f_{1}^{t\rightarrow Wb'}$ as \begin{eqnarray}
f_{1}^{t\rightarrow Wb'({\rm real})} & = & \frac{\alpha_{s}}{4\pi}C_{F}C_{\epsilon}\left\{ \frac{1}{\epsilon^{2}}+\frac{5}{2\epsilon}-\frac{2}{\epsilon}\ln(1-\beta_{W})-\frac{2\pi^{2}}{3}+\ln^{2}2-2\ln2\right.\nonumber \\
 &  & -2{\textrm{Li}}_{2}(\frac{1-\beta_{W}}{2-\beta_{W}})-\frac{7}{2}\ln\frac{s_{min}}{m_{t}^{2}}+2\ln2\ln(1-\beta_{W})\nonumber \\
 &  & +\frac{4-2\beta_{W}}{1-\beta_{W}}\ln(2-\beta_{W})-\ln^{2}(2-\beta_{W})-\ln^{2}(1-\beta_{W})\nonumber \\
 &  & \left.-\ln^{2}\frac{s_{min}}{m_{t}^{2}}+4\ln(1-\beta_{W})\ln\frac{s_{min}}{m_{t}^{2}}-\frac{1}{1-\beta_{W}}+I_{{\rm \, Scheme}}^{\, t\rightarrow Wb'({\rm real})}\right\} ,\label{eq:I-real-tdec}\end{eqnarray}
where the scheme dependent term $I_{\,{\rm Scheme}}^{\, t\rightarrow Wb'({\rm real})}$
is\begin{equation}
I_{\,{\rm Scheme}}^{\, t\rightarrow Wb'({\rm real})}=\begin{cases}
{\displaystyle \frac{7}{2}} & {\rm in~DREG~scheme},\\
3 & {\rm in~DRED~scheme}.\end{cases}\end{equation}

The resolved region of $t\rightarrow Wb'g$ is obtained by multiplying
the following function to the phase space: \begin{equation}
\biggl[\Theta(\left|s_{tg}\right|+\left|s_{b'g}\right|-2s_{min})-\Theta(\left|s_{tg}\right|-2s_{min})\Theta(s_{min}-\left|s_{b'g}\right|)\biggr].\label{eq:pss-condition-tdec-2}\end{equation}
 We have checked the formulas of (\ref{decay1})-(\ref{eq:pss-condition-tdec-2})
by comparing the result of NLO correction to $\Gamma(t\rightarrow Wb')$
with Ref.~\cite{Li:1990qf}.

\section{Combining the production and decay processes}

With those building blocks given in the above sections, the NLO QCD
corrections to single top quark production and decay can be computed,
keeping the full information on the spin configuration of the intermediate
top quark state. The general differential hadronic cross section at
NLO can be written as 
\begin{eqnarray}
 &  & d\sigma(H_{1}H_{2}\rightarrow YX)\nonumber \\
 & = & \sum\limits _{a,b}\int dx_{1}dx_{2}\Biggl\{ f_{a}^{H_{1}}(x_{1},\mu_{F})f_{b}^{H_{2}}(x_{2},\mu_{F})\times[d\sigma_{0}(ab\rightarrow Y)+d\sigma_{1}(ab\rightarrow Y)]\nonumber \\
 &  & \,\,\,\,\,\,\,\,\,\,\,\,\,\,\,\,\,\,\,\,\,\,\,\,\,\,\,\,\,\,\,\,\,\,\,\,+\alpha_{s}f_{a}^{H_{1}}(x_{1},\mu_{F})C_{b}^{H_{2}}(x_{2},\mu_{F},s_{min})d\sigma_{0}(ab\rightarrow Y)\label{eq:nlo_cs}\nonumber \\
 &  & \,\,\,\,\,\,\,\,\,\,\,\,\,\,\,\,\,\,\,\,\,\,\,\,\,\,\,\,\,\,\,\,\,\,\,\,+\alpha_{s}C_{a}^{H_{1}}(x_{1},\mu_{F},s_{min})f_{b}^{H_{2}}(x_{2},\mu_{F})d\sigma_{0}(ab\rightarrow Y)+(x_{1}\leftrightarrow x_{2})\Biggr\},
\end{eqnarray}
where $d\sigma_{0}$ is the leading-order subprocess cross section,
$d\sigma_{1}$ is the $O(\alpha_{S})$ ``crossed" subprocess cross section, cf. Fig.~\ref{fig:crossing}.

We now consider the single top quark production subprocess $ab\rightarrow t_{\lambda}h_{1}$
with $t_{\lambda}\rightarrow W_{\rho}h_{2}$ and $W_{\rho}\rightarrow l\nu$.
(Here, $h_{1}$ and $h_{2}$ stand for any single parton or multiple
partons. $\lambda$ and $\rho$ are the top quark spin and $W$ boson
polarization indices, respectively) In the frame work of NWA, the cross section
can be written as 
\begin{eqnarray}
d\sigma(ab\rightarrow l\nu h_{1}h_{2}) & = & \frac{1}{2\hat{s}}|\sum\limits _{\lambda,\rho}M(ab\rightarrow t_{\lambda}h_{1})M(t_{\lambda}\rightarrow W_{\rho}h_{2})M(W_{\rho}\rightarrow l\nu)|^{2}\label{subcross}\nonumber \\
 & \times  & S_{F}\frac{1}{2m_{t}\Gamma_{t}}\frac{1}{2m_{W}\Gamma_{W}}d\Phi(ab\rightarrow th_{1})d\Phi(t\rightarrow Wh_{2})d\Phi(W\rightarrow\ell\nu),
\end{eqnarray}
where $S_{F}$ denotes the proper spin and color factors, $d\Phi$'s
are the phase space elements (${\displaystyle (2\pi)^{4}\delta^{4}(P-\sum p_{i})\prod\frac{d^{3}\vec{p}_{i}}{2E_{i}(2\pi)^{3}}}$).
At the LO, $\Gamma_t$ in the above equation should be replaced by the Born level decay width $\Gamma^{0}_t(t\rightarrow bW)$. At the
NLO, special cares should be taken to assign $\Gamma_t$, cf. Eq.~(\ref{eq:nlo_hardpart}). $\Gamma_{W}$ is the $W$ boson total decay width.

The matrix element square can be calculated as follows. The sum over
$\rho$ (the polarization state of the $W$ boson from top decay) is equivalent to the following replacement in $M(t_{\lambda}\rightarrow W_{\rho}h_{2})$:
\begin{eqnarray}
 &  & \varepsilon_{W}^{\mu}\rightarrow\frac{g}{2\sqrt{2}}\bar{u}_{\nu}\gamma^{\mu}(1-\gamma_{5})v_{\ell}.\label{wpol}\end{eqnarray}
 We denote the result by $M(t_{\lambda}\rightarrow Wh_{2})$. Decomposing
$M(ab\rightarrow t_{\lambda}h_{1})$ and $M(t_{\lambda}\rightarrow Wh_{2})$ by 
\begin{eqnarray}
M(ab\rightarrow t_{\lambda}h_{1}) & = & \bar{u}_{\lambda}(p_{t})M^{prd},\\
M(t_{\lambda}\rightarrow Wh_{2}) & = & M^{dec}u_{\lambda}(p_{t}), 
\end{eqnarray}
where we have explicitly separated the on-shell top quark spinors from both the production
and the decay matrix elements, then we have \begin{eqnarray}
|\sum\limits _{\lambda,\rho}M(ab\rightarrow t_{\lambda}h_{1})M(t_{\lambda}\rightarrow W_{\rho}h_{2})M(W_{\rho}\rightarrow l\nu)|^{2}=|M^{dec}(\rlap/p_{t}+m_{t})M^{prd}|^{2}.\label{decom}\end{eqnarray}
 In our calculations, $M^{dec}$ and $M^{prd}$ are calculated numerically
using helicity amplitude approach and can be easily obtained from
the formulas presented in the sections \ref{sec:Born-level-matrix}
and \ref{sec:NLO-matrix-element}. Eqs.~(\ref{wpol}) and (\ref{decom})
guarantee that the spin and angular correlations of the decay products
are preserved.

Denoting \begin{eqnarray}
d\Phi^{{\rm LO}} & = & S_{F}\frac{1}{2m_{t}\Gamma_{t}^{0}}\frac{1}{2m_{W}\Gamma_{W}}d\Phi(ab\rightarrow th_{1})d\Phi(t\rightarrow Wh_{2})d\Phi(W\rightarrow\ell\nu),\\
d\Phi^{{\rm NLO}} & = & S_{F}\frac{1}{2m_{t}\Gamma_{t}}\frac{1}{2m_{W}\Gamma_{W}}d\Phi(ab\rightarrow th_{1})d\Phi(t\rightarrow Wh_{2})d\Phi(W\rightarrow\ell\nu),\end{eqnarray}
where $\Gamma_{t}=\Gamma_{t}^{0}(t\rightarrow bW)+\Gamma_{t}^{1}(t\rightarrow bW)$
and $\Gamma_{t}^{1}(t\rightarrow bW)$ is the $O(\alpha_{S})$ correction
to the Born level decay width $\Gamma_{t}^{0}(t\rightarrow bW)$, the LO subprocess cross section is \begin{eqnarray}
d\sigma_{0}(ab\rightarrow l\nu h_{1}h_{2}) & = & \frac{1}{2\hat{s}}|M_{0}^{dec}(\rlap/p_{t}+m_{t})M_{0}^{prd}|^{2}d\Phi^{{\rm LO}},\end{eqnarray}
where $M_{0}^{prd,dec}$ stand for the LO amplitude. The NLO "crossed" subprocess
cross section is \begin{eqnarray}
d\sigma_{1}(ab\rightarrow l\nu h_{1}h_{2}) & = & \frac{1}{2\hat{s}}|M_{0}^{dec}(\rlap/p_{t}+m_{t})M_{1R}^{prd}|^{2}d\Phi^{{\rm LO}}\nonumber \\
 & + & \frac{1}{2\hat{s}}2{\textrm{Re}}\big[M_{0}^{dec}(\rlap/p_{t}+m_{t})M_{1SCV}^{prd}(M_{0}^{dec}(\rlap/p_{t}+m_{t})M_{0}^{prd})^{\dagger}\big]d\Phi^{{\rm LO}}\nonumber \\
 & + & \frac{1}{2\hat{s}}2{\textrm{Re}}\big[M_{1SCV}^{dec}(\rlap/p_{t}+m_{t})M_{0}^{prd}(M_{0}^{dec}(\rlap/p_{t}+m_{t})M_{0}^{prd})^{\dagger}\big]d\Phi^{{\rm NLO}}\nonumber \\
 & + & \frac{1}{2\hat{s}}|M_{1R}^{dec}(\rlap/p_{t}+m_{t})M_{0}^{prd}|^{2}d\Phi^{{\rm NLO}},\label{eq:nlo_hardpart}\end{eqnarray}
where $M_{1SCV,1R}^{prd,dec}$ stand for the $O(\alpha_{S})$ amplitudes contributed
from either soft+collinear+virtual or resolved real corrections for the production
or decay processes. The first term is the real NLO correction from
production. The second term is the soft+collinear+virtual correction
from production. The last two terms are the corrections from the top quark decay.
If no kinematical cut is applied, the last two terms cancel each other,
which means there is no net correction to the cross section from the
top quark decay. Because the virtual correction processes and the real
correction processes have different phase spaces $d\Phi^{{\rm LO}}$
and $d\Phi^{{\rm NLO}}$, we calculate them separately using different
Monte Carlo programs.

\section{Conclusions}

Precision measurement of the single top quark events requires more
accurate theoretical prediction. The fully NLO differential cross
section for on-shell single top quark production has been calculated
two years ago, but NLO corrections to the top quark decay process
are not included, nor the effect of the top quark width. Since the
top quark production and decay do not occur in isolation from each
other, a theoretical study that includes both kinds of corrections
and keeps the spin correlations between the final state particles
is in order. 

In this paper we have presented a complete calculation of NLO QCD
corrections to  both s-channel and t-channel single top quark production
and decay processes at hadron colliders. In our calculation the phase
space slicing method with one cutoff scale is adopted because it takes
advantage of the generalized crossing property of the NLO matrix elements
to reduce the analytical calculations. After calculating the effective
form factors with all the partons in the final state, we can easily
cross the needed partons into the initial state to calculate the s-channel
or t-channel single top quark cross sections. To respect the spin correlations between the
final state particles, all the amplitudes are calculated using the
helicity amplitude method. The form factor approach is used for including
the SCV (soft+collinear+virtual) corrections so that our results can also be used to study
new physics effects that result in the similar form factors. To consider
the top quark production with top quark decay consistently, the 
``modified narrow
width approximation'', cf. Sec.~\ref{sub:nwa}, is adopted in our calculation. 
Our results are given in both the DREG ('t Hooft-Veltman $\gamma_5$)~\cite{'tHooft:1972fi} 
and DRED~\cite{Bern:2002zk} schemes
to treat the $\gamma_5$ matrix in the scattering amplitudes which is important for predicting the
distributions of final state particles.

A preliminary study on the
phenomenology of single top physics at Tevatron collider based on the 
theoretical framework presented in this paper was already presented in
Ref.~\cite{caotalk}.
A more detailed study on the phenomenology
predicted by our calculations will be presented in sequential paper~\cite{pheno-schan}.

\emph{Notes added}: While completing the writing of this paper, we
noted that another article dealing with the same subject, but with
different method, just appeared~\cite{Campbell:2004up}.

\begin{acknowledgments}
We thank Hong-Yi Zhou for collaboration in the early stage
of this project, and F. Larios for a critical reading of the manuscript.
CPY thanks the hospitality of National Center for
Theoretical Sciences in Taiwan, ROC, where part of this work was completed.
This work was supported in part by the NSF grants
PHY-0244919 and PHY-0100677.
\end{acknowledgments}

\newpage

\appendix

\section{Helicity Amplitudes\label{sec:Helicity-notation}}

In this appendix we briefly summarize our method for calculating the
helicity amplitudes. The method breaks down the algebra of four-dimensional
Dirac spinors and matrices into equivalent two-dimensional ones. In
the Weyl basis, Dirac spinors have the form\begin{equation}
\left(\begin{array}{c}
\psi_{+}\\
\psi_{-}\end{array}\right),\end{equation}
where for fermions
\begin{equation}
\psi_{\pm}=\biggl\{\begin{array}{cc}
u_{\pm}^{(\lambda=1)}=\omega_{\pm}\chi_{1/2}\\
u_{\pm}^{(\lambda=-)}=\omega_{\mp}\chi_{-1/2} & ,\end{array}
\end{equation}
and anti-fermions
\begin{equation}
\psi_{\pm}=\biggl\{\begin{array}{cc}
v_{\pm}^{(\lambda=1)}=\pm\omega_{\mp}\chi_{-1/2}\\
v_{\pm}^{(\lambda=-)}=\mp\omega_{\pm}\chi_{1/2} & ,\end{array}
\end{equation}
with $\omega_{\pm}=\sqrt{E\pm\left|\vec{p}\right|}$, where $E$ and
$\vec{p}$ are the energy and momentum of the fermion, respectively.
Explicitly, in spherical coordinates,
\begin{equation}
p^{\mu}=\left(E,~\left|\vec{p}\,\right|\sin\theta\cos\phi,\left|\vec{p}\,\right|\sin\theta\sin\phi,
\left|\vec{p}\,\right|\cos\theta\right)
\end{equation}
The $\chi_{\lambda/2}$'s are eigenvectors of the helicity operator
\begin{equation}
h=\hat{p}\cdot\sigma,
\end{equation}
where $\hat{p}=\vec{p}/\left|\vec{p}\right|$ and the 
eigenvalue $\lambda=1$ stands for ``spin-up'' fermion and $\lambda=-1$
for ``spin-down'' fermion.\begin{equation}
\chi_{1/2}\equiv|\hat{p}+\!\!\!>=\left(\begin{array}{c}
\cos\theta/2\\
e^{i\phi}\sin\theta/2\end{array}\right),\,\,\chi_{-1/2}\equiv|\hat{p}-\!\!\!>=\left(\begin{array}{c}
-e^{i\phi}\sin\theta/2\\
\cos\theta/2\end{array}\right),\label{eq:braket}\end{equation}
where we introduce the shorthand notations $|\hat{p}\pm\!\!>$ for
$\chi_{\pm1/2}$. Furthermore, 
\begin{equation}
<\!\hat{p}\pm|=\left(|\hat{p}\pm\!\!>\right)^{\dagger}
\end{equation}
where the superscript denotes taking hermitian complex conjugation. Under the operation of charge conjugation,
denoted as $|\widetilde{\hat{p}+}\!>$, we have
\begin{equation}
|\widetilde{\hat{p}+}\!>\equiv i\sigma_2~|\hat{p}+\!>^{*}= -|\hat{p}-\!>.
\end{equation}
Similarly,
\begin{eqnarray}
  |\widetilde{\hat{p}-}\!>&=&+~|~\hat{p}+\!>,\\
  <\widetilde{\hat{p}+}|&=&-<\hat{p}-\!|,\\
  <\widetilde{\hat{p}-}|&=&+<\hat{p}+\!|.
\end{eqnarray}

Gamma matrices in the Weyl basis have the form\begin{equation}
\gamma^{0}=\left(\begin{array}{cc}
0 & 1\\
1 & 0\end{array}\right),\,\,\,\gamma^{j}=\left(\begin{array}{cc}
0 & -\sigma_{j}\\
\sigma_{j} & 0\end{array}\right),\,\,\,\gamma^{5}=\left(\begin{array}{cc}
1 & 0\\
0 & -1\end{array}\right),\end{equation}
where $\sigma_{j}$ are the Pauli $2\times2$ spin matrices. In the
Weyl basis, $\not\! p$ takes the form\begin{equation}
\not\! p\equiv p_{\mu}\gamma^{\mu}=\left(\begin{array}{cc}
0 & p_{0}+\vec{\sigma}\cdot\vec{p}\\
p_{0}-\vec{\sigma}\cdot\vec{p} & 0\end{array}\right)\equiv\left(\begin{array}{cc}
0 & \not\! p_{+}\\
\not\! p_{-} & 0\end{array}\right)\equiv p_{\mu}\left(\begin{array}{cc}
0 & \gamma_{+}^{\mu}\\
\gamma_{-}^{\mu} & 0\end{array}\right)\end{equation}
where \begin{equation}
\gamma_{\pm}^{\mu}=(1,\mp\vec{\sigma}).\end{equation}

\section{$A_{h}$ and $B_{h}$ in the Crossing Functions\label{sec:Crossing-Functions}}

There are four independent $A_{p\rightarrow hX}(x,\mu_{F})$ and $B_{p\rightarrow hX}^{{\rm Scheme}}(x,\mu_{F})$ coefficient functions
in the process independent, but $s_{min}$ and factorization scheme
dependent crossing functions, cf. Eq.~(\ref{eq:crossing_def}). They are listed below, after suppressing the $\mu_F$ dependence.
\begin{eqnarray}
A_{g\rightarrow gg}(x) & = & [\frac{33-2n_{f}}{18}+2\ln(1-x)]f_{g}^{H}(x)\\
 &  & +2\int_{x}^{1}dzf_{g}^{H}(x/z)[\frac{1-z}{z^{2}}+1-z]+2\int_{x}^{1}dz\frac{f_{g}^{H}(x/z)-f_{g}^{H}(x)}{1-z},\nonumber \\
A_{q\rightarrow qg}(x) & = & [\frac{2}{3}+\frac{8}{9}\ln(1-x)]f_{q}^{H}(x)+\frac{4}{9}\int_{x}^{1}dz\frac{(1+z^{2})/zf_{q}^{H}(x/z)-2f_{q}^{H}(x)}{1-z},\\
A_{g\rightarrow q\bar{q}}(x) & = & \frac{1}{6}\int_{x}^{1}dzf_{g}^{H}(x/z)\frac{z^{2}+(1-z)^{2}}{z},\\
A_{q\rightarrow gq}(x) & = & \frac{4}{9}\int_{x}^{1}dzf_{q}^{H}(x/z)\frac{1+(1-z)^{2}}{z^{2}},\end{eqnarray}
\begin{eqnarray}
B_{g\rightarrow gg}^{\overline{MS}}(x) & = & [\frac{\pi^{2}}{3}-\frac{67}{18}+\frac{5n_{f}}{27}+\ln^{2}(1-x)]f_{g}^{H}(x)\\
 &  & +2\int_{x}^{1}dzf_{g}^{H}(x/z)\ln(1-z)[\frac{1-z}{z^{2}}+1-z]\\
 &  & +2\int_{x}^{1}dz\ln(1-z)\frac{f_{g}^{H}(x/z)-f_{g}^{H}(x)}{1-z},\\
B_{q\rightarrow qg}^{\overline{MS}}(x) & = & [\frac{8}{9}(\frac{\pi^{2}}{6}-\frac{7}{4})+\frac{4}{9}\ln^{2}(1-x)]f_{q}^{H}(x)+\frac{4}{9}\int_{x}^{1}dzf_{q}^{H}(x/z)\frac{1-z}{z}\\
 &  & +\frac{4}{9}\int_{x}^{1}dz\ln(1-z)\frac{(1+z^{2})/zf_{q}^{H}(x/z)-2f_{q}^{H}(x)}{1-z},\\
B_{g\rightarrow q\bar{q}}^{\overline{MS}}(x) & = & \frac{1}{6}\int_{x}^{1}dzf_{g}^{H}(x/z)[\frac{z^{2}+(1-z)^{2}}{z}\ln(1-z)+2(1-z)],\\
B_{q\rightarrow gq}^{\overline{MS}}(x) & = & \frac{4}{9}\int_{x}^{1}dzf_{q}^{H}(x/z)[\frac{1+(1-z)^{2}}{z^{2}}\ln(1-z)+1],\end{eqnarray}
\begin{eqnarray}
B_{g\rightarrow gg}^{{\rm DRED}}(x) & = & \frac{\pi^{2}}{3}-\frac{67}{18}+\frac{5n_{f}}{27}+\ln^{2}(1-x)]f_{g}^{H}(x)\\
 &  & +2\int_{x}^{1}dzf_{g}^{H}(x/z)\ln(1-z)[\frac{1-z}{z^{2}}+1-z]\\
 &  & +2\int_{x}^{1}dz\ln(1-z)\frac{f_{g}^{H}(x/z)-f_{g}^{H}(x)}{1-z},\\
B_{q\rightarrow qg}^{{\rm DRED}}(x) & = & \frac{8}{9}(\frac{\pi^{2}}{6}-\frac{3}{2})+\frac{4}{9}\ln^{2}(1-x)]f_{q}^{H}(x)-\frac{4}{9}\int_{x}^{1}dzf_{q}^{H}(x/z)\frac{1-z}{z}\\
 &  & +\frac{4}{9}\int_{x}^{1}dz\ln(1-z)\frac{(1+z^{2})/zf_{q}^{H}(x/z)-2f_{q}^{H}(x)}{1-z},\\
B_{g\rightarrow q\bar{q}}^{{\rm DRED}}(x) & = & \frac{1}{6}\int_{x}^{1}dzf_{g}^{H}(x/z)[\frac{z^{2}+(1-z)^{2}}{z}\ln(1-z)-2(1-z)],\\
B_{q\rightarrow gq}^{{\rm DRED}}(x) & = & \frac{4}{9}\int_{x}^{1}dzf_{q}^{H}(x/z)[\frac{1+(1-z)^{2}}{z^{2}}\ln(1-z)+1],\end{eqnarray}
where $n_{f}$ is the flavor number, $f_{h}^{H}(x)$ is the parton
distribution function of parton $h$ inside hadron $H$. In the above,
we have set $N_{c}=3$. The subscript $\overline{MS}$ indicates the
results in the $\overline{MS}$ DREG scheme while the subscript DRED
indicates the results in the DRED scheme.
\end{document}